\documentclass[12pt]{article}  
\usepackage{a4}
\usepackage{graphicx}
\usepackage{subfigure}
\usepackage[centertags]{amsmath}
\usepackage{amssymb}

\input diagrams

\newcommand{\unity}{{ 1\:\!\!\!\mbox{I}}}

\newcommand{\Z}{{\mathbb Z}}
\newcommand{\OR}{\Omega {\mathcal R}}
\newcommand{\A}{\mbox{\bf \scriptsize A}}
\newcommand{\B}{\mbox{\bf \scriptsize B}}

\newcommand{\AN}{{\mathcal A}}
\newcommand{\MS}{{\mathcal M}}
\newcommand{\CP}[6]{\text{tr} \left[ 
    \gamma^{(#1,#2)-T}_{\OR #5 #6} 
    \gamma^{(#3,#4)}_{\OR #5 #6} \right ]}
\newcommand{\LAT}{{\mathcal L}}
\newcommand{\KB}{{\mathcal K}}

\newcommand{\barre}[1]{%
        \setbox1=\hbox{$#1$} \dimen2=\ht1 \dimen3=\dp1 \dimen4=\wd1  
        \setbox2=\hbox{\sl /}  
        \dimen1=\wd1 \advance\dimen1 by -\wd2 \divide\dimen1 by 2  
        \advance\dimen1 by \wd2 \advance\dimen1 by 0.4pt  
        \setbox3=\hbox to \wd1{\hss \box1 \kern -\dimen1 \box2\hss}  
        \ht3=\dimen2 \dp3=\dimen3 \wd3=\dimen4  
        \box3  
        }  
\begin{document}  
\pagestyle{empty}  
\begin{flushright}  
                     hep-th/0008250 
\end{flushright}  
\vskip 2cm    

\begin{center}  
{\huge Supersymmetric $\Z_N\times \Z_M$ Orientifolds
    in 4D with D-Branes at Angles\\}  
\vspace*{5mm} \vspace*{1cm}   
\end{center}  
\vspace*{5mm} \noindent  
\vskip 0.5cm  
\centerline{\bf Stefan F\"orste, Gabriele Honecker and
Ralph Schreyer}
\vskip 1cm
\centerline{\em Physikalisches Institut, Universit\"at Bonn}
\centerline{\em Nussallee 12, D-53115 Bonn, Germany}
\vskip2cm
  
\centerline{\bf Abstract}  

We construct orientifolds of type IIA string theory. The theory is
compactified on a \linebreak  $T^6 /\Z_N\times \Z_M$ orbifold.
In addition worldsheet parity in combination with a reflection of
three compact directions is modded out. Tadpole cancellation requires
to add D-6-branes at angles. The resulting four dimensional theories are
${\cal N}=1$ supersymmetric and non-chiral.

\vskip .3cm

  
\newpage  

\setcounter{page}{1} \pagestyle{plain}  

\section{Introduction} \label{intro}

One of the major issues in string theory is to classify consistent
theories in especially 3+1 dimensions. Insights into strong
coupling regions of string theory provide reasons to hope that
apparently different models are actually equivalent and can be mapped onto
each other by duality transformations. Often, strong and weak coupling
regions are interchanged in this process. Open strings with Dirichlet
boundary conditions e.g.\ provide a perturbative description of
solitonic (non-perturbative) objects (D-p-branes) in type II string
theories\cite{Polchinski:1995mt}. This observation was crucial for one
of the first conjectures about string dualities - the heterotic/type I
duality\cite{Polchinski:1996df}. Since compactifications of the
heterotic string are of particular phenomenological interest one
expects the same for type I compactifications. Here, one typically
starts with a type II theory on an orbifold. In addition, worldsheet
parity (possibly combined with discrete targetspace transformations)
is modded out. The consistency requirement of modular invariance is
replaced by tadpole cancellation conditions. The resulting models are
called orientifolds. This kind of construction has been considered
already some time
ago[\ref{ago1}--\ref{agol}].
The first 
formulation using the modern
language of D-branes and Orientifold fixed planes
was given in\cite{Gimon:1996rq}.
Subsequently, several models with different numbers of non-compact
dimensions and unbroken supersymmetries have been constructed
e.g. in[\ref{or1}--\ref{orl}].

In orientifolds the amount of unbroken supersymmetry depends on the
orbifold group and the arrangement of D-branes and O-planes needed for
consistent compactifications. In\cite{Berkooz:1996km} it was pointed
out that also D-branes intersecting at angles can leave some
supersymmetries unbroken. A concrete realization of this observation
in orientifold constructions was worked out in
\cite{Blumenhagen:2000md,Blumenhagen:2000ev}. These authors considered
type IIB (IIA) string theory on $T^4/\Z_N$ ($T^6/\Z_N$) orbifolds with
$N=3,4,6$. In addition, they gauged worldsheet parity together with
the reflection of two (three) directions of the $T^4/\Z_N$
($T^6/\Z_N$) orbifold 
such that this reflection leaves O-planes intersecting at angles
fixed. To cancel the corresponding RR-charges, D-branes intersecting at
angles need to be added. In the present paper we are going to
supplement this class of models by compactifying type IIA on a 
$T^6/\left(\Z_N\times \Z_M\right)$ orbifold together with imposing
invariance under 
worldsheet parity inversion combined with the reflection of three orbifold
directions. We will discuss only models with ${\cal N}=1$ supersymmetry in
four dimensions. All possible compactifications of this kind yield
non-chiral four dimensional models with different gauge groups and
matter content.

In the next section we describe general features of the
construction. The third section is devoted to a detailed study of the
$\Z_4 \times \Z_2$ orientifold. Subsequently, we briefly give results
for all other consistent models, {\it viz.}\ the $\Z_2\times \Z_2$,
$\Z_6\times \Z_3$ and the $\Z_3\times \Z_3$ orientifolds. We conclude
by summarizing our results. Three appendices provide technical details
of the considered models: appendix A is addressed to the computation
of loop channel diagrams, appendix B contains tables of massless
spectra, and appendix C describes a projective representation used for
the $\Z_4\times \Z_2$ orientifold.


\section{General Setup}\label{setup}

Throughout the paper we will discuss models with four non-compact
dimensions labeled by $x^\mu$, $\mu = 0, \ldots ,3$. In addition, there
are six compact directions which we describe by three complex
coordinates,
\begin{equation}
z^{1}=x^4 + ix^5 \, ,\; z^2 =x^6+ix^7 \, ,\,\; z^3 = x^8+ix^9 .
\end{equation}
Each of those coordinates describes a torus $T^2$. In addition, 
points on these
tori are identified under rotations 
\begin{equation} \label{orbifoldaction}
\Theta_1 :\; z^j \rightarrow e^{2\pi iv_j}z^j \; ,\;\; \Theta_2:\; z^j
\rightarrow e^{2\pi iw_j}z^j ,
\end{equation}
where $\left(\Theta_1 , \Theta_2\right)$ denotes an element of the
orbifold group 
$\Z_N\times \Z_M$. (The action on the complex conjugated coordinates
just follows from the conjugation of (\ref{orbifoldaction}).)
We will discuss type IIA theories on these manifolds. In addition, we
gauge the symmetry generated by $\Omega {\cal R}$ where $\Omega$
reverses worldsheet parity, and ${\cal R}$ reflects the imaginary
parts of the $z^i$,
\begin{equation}
{\cal R}: z^i \rightarrow \bar{z}^i .
\end{equation}   
The gauging of $\Omega {\cal R}$ creates Orientifold fixed planes
(O-planes). The location of those planes is given by sets of 
points fixed under the elements of $\Omega {\cal R}\times \Z_N\times
\Z_M$. These 
fixed planes carry RR-charges which must be canceled by adding
D-branes to the model\cite{Polchinski:1995mt,Gimon:1996rq}.
One of these O-planes is extended along the non-compact directions
and the real parts of the $z^i$. It is invariant under the ${\cal R}$
reflection. To visualize the remaining O-planes we note that ($I=1,2$) 
\begin{equation} \label{durchziehen}
\Omega {\cal R} \Theta_I: z^i \; =\; \left(\Theta_I\right)^{-\frac{1}{2}}
\Omega {\cal R} \left(\Theta_I\right)^{\frac{1}{2}}: z^i
\end{equation}
where the powers of $\pm \frac{1}{2}$ indicate that the rotations are
performed with plus-minus half the angle as compared to
(\ref{orbifoldaction}). 
Thus, a fixed plane under $\Omega {\cal R}\Theta_I$ is obtained by acting with
$\left(\Theta_I\right)^{-\frac{1}{2}}$  on the set of points fixed 
under ${\cal R}$.
Therefore (starting from the O-6-plane at $z^i
=\bar{z}^i$, $i=1,2,3$), one obtains a set of O-6-planes
intersecting at angles, whose values are given by half the order of
the corresponding $\Z_N\times \Z_M$ element. 
Since all these O-6-planes have compact
transverse directions and carry RR-charges, one expects that for
consistency one needs to include a certain amount of D-6-branes canceling
exactly those charges. 
These D-6-branes need to be parallel to the O-6-planes and hence also
intersect at angles.
Indeed, for $\Z_N$ orbifolds this has been shown
to be the case\cite{Blumenhagen:2000md,Blumenhagen:2000ev}.
In the following sections we will generalize those models to
$\Z_N\times \Z_M$ orbifolds. This turns out to be a straightforward
modification of the discussion given
in\cite{Blumenhagen:2000md,Blumenhagen:2000ev}. A new ingredient,
however, is that in some cases more complicated projective
representations of the orientifold group in the open string sector are
needed. This has been observed before in some $\Z_N\times \Z_M$
orientifolds of type IIB models\cite{Berkooz:1997dw,Forste:1998bd}. In
fact, for the $\Z_2\times \Z_2$ model to be discussed in section
\ref{zweikreuzzwei} we will obtain the T-dual version of the model
of\cite{Berkooz:1997dw}. 

In the next paragraphs we are going to review some of the technical
aspects necessary for orientifold constructions. The main consistency
requirement comes from RR charge conservation. Technically, it
translates into the tadpole cancellation condition\cite{Gimon:1996rq}.
The RR charges describe the size of the couplings of  O-planes and
D-branes to RR gauge fields. The numerical value of these couplings
can be computed by extracting the RR exchange contribution to the
forces acting between O-planes and D-branes\cite{Polchinski:1995mt}.
A convenient way of computing these forces is to move to the
open string loop channel. 
There, the RR charge of O-planes and D-branes can be extracted from
the UV-limits of
the following parts of the Klein bottle, M\"obius strip and annulus
diagrams\cite{Polchinski:1988tu}:
\begin{equation} \label{closedopen}
\begin{array}{r l}
\mbox{Klein bottle:} & \mbox{Closed string  NSNS states with {\bf
    P}$\Omega{\cal R}\left(-\right)^F$ insertion}\\
\mbox{M\"obius strip:} & \mbox{Open string R states with $-${\bf
    P}$\Omega{\cal R}$ insertion} \\
\mbox{Annulus:} & \mbox{Open string NS states with {\bf
    P}$\left(-\right) ^F$ insertion}
\end{array}
\end{equation}
Here, $\left( -\right)^F$ is the fermion number to be defined
below. (For closed strings $\left( -\right)^F = \left( -\right)^{F_L}=
  \left(-\right)^{F_R}$ because of the presence of $\Omega$ in the trace.)
    Further, we denote by {\bf P} the projector on  
states invariant under the orbifold group $\Z_N\times \Z_M$. 
The requirement of tadpole cancellation determines the number of
D-6-branes and part of the representation of the orientifold group on
the Chan Paton indices.

\begin{figure}
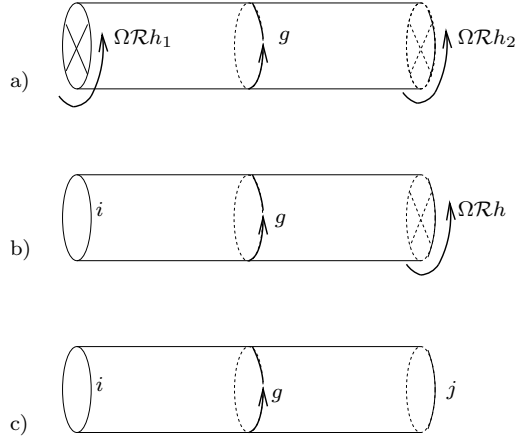
  
\begin{center}
\input diagrams.pstex_t
\end{center}
\caption{a) Klein bottle, b) M\"obius strip, c) Cylinder}
\label{diagrams}
\end{figure}

Another essential consistency condition is what is called
``completion of the projector in the tree channel''
in\cite{Blumenhagen:2000md}. Let us briefly recall their
arguments. The important diagrams are drawn in figure
\ref{diagrams}. The notation is taken
from\cite{Gimon:1996rq}. Orbifold group elements are denoted by $h$ or
$g$. In the tree channel picture the crosscaps correspond to O-planes
invariant under $\Omega {\cal R} h$. D-branes are assigned a letter
$i$ or $j$. The orbifold group element $g$ denotes the twist sector of
the closed strings propagating in the tree channel. (For further
details see\cite{Gimon:1996rq}.) Consistency of the boundary
conditions requires
\begin{equation}\label{cona}
\left( \Omega {\cal R} h_1\right)^2 = \left( \Omega {\cal R}
  h_2\right)^2 =g
\end{equation}
in the Klein bottle diagram and 
\begin{equation}\label{conb}
\left( \Omega {\cal R} h\right)^2 =g
\end{equation}
in the M\"obius strip. For the class of orientifolds discussed here,
the lhs of (\ref{cona}) and (\ref{conb}) are the identity. Hence, in
the tree channel only untwisted closed strings propagate in the Klein
bottle and in the M\"obius strip. Since these must be invariant under
the orbifold group, the tree channel amplitude must contain the
insertion of the complete projector on invariant states. 
The actual calculation of the diagrams depicted in figure
\ref{diagrams} will be done in the loop channel, where worldsheet time
is vertical. Transforming back to the tree channel one must recover
the insertion of the complete projector mentioned above. For the Klein
bottle amplitude this requirement leads to restrictions on the
possible orbifold lattices. In the M\"obius strip one obtains further
conditions for the representation of the orientifold group on the
Chan-Paton matrices. 

Finally, let us introduce some notation and
definitions which we
mainly borrow from\cite{Blumenhagen:2000ev}. The action of the
orientifold group on closed string degenerate
ground states\footnote{All states are in light cone gauge. The first
  entry in the state corresponds to two  
  non-compact directions, whereas the other three belong to the three
  complex compact directions.} is specified by
\begin{eqnarray}
{\cal R}: \left| s_0, s_1, s_2, s_3\right> & \rightarrow & \left| s_0,
  -s_1, -s_2, -s_3\right> \nonumber \\
\Theta_1:  \left| s_0, s_1, s_2, s_3\right>& \rightarrow & e^{2\pi i
  \vec{v}\cdot \vec{s}}\left| s_0, s_1, s_2, s_3\right> \label{phases}\\
\Theta_2:  \left| s_0, s_1, s_2, s_3\right>& \rightarrow & e^{2\pi i
  \vec{w}\cdot \vec{s}}\left| s_0, s_1, s_2, s_3\right> \nonumber
\end{eqnarray}
Worldsheet parity inversion $\Omega$ interchanges the left with the
right moving sector.
Under GSO projection states with fermion numbers $\left(
  -1\right)^{F_L}= \left(-1\right)^{F_R}= 1$ are kept, where
\begin{eqnarray}
\left( -1\right)^{F_L} \left| s_0, s_1, s_2, s_3\right> & = & -
e^{\pi i\left( s_0 - s_1 - s_2 - s_3\right)}\left| s_0, s_1, s_2,
    s_3\right>\nonumber \\
\left( -1\right)^{F_R} \left| s_0, s_1, s_2, s_3\right> & = &
-e^{\pi i\left( s_0 + s_1 + s_2 + s_3\right)}\left| s_0, s_1, s_2,
    s_3\right>\label{gso}
\end{eqnarray}
In the loop channel the expressions for the Klein bottle, M\"obius
strip and annulus are ($c =V_4/\left(8\pi \alpha^\prime\right)^2$
  and $V_4$  is the regularized volume of non-compact momentum space)
\begin{eqnarray}
{\cal K} & = & 4c\int_0^\infty \frac{dt}{t^3} \mbox{Tr}_{U+T}\left(
  \frac{\Omega {\cal R}}{2}\mbox{\bf P}\left(\frac{1
  +\left(-1\right)^{F}}{2}\right)\left(-1\right)^{\mbox{\scriptsize \bf S}} 
e^{-2\pi t\left( L_0+\bar{L}_0\right)}\right)
  , \label{kb} \\
{\cal A} & = & c \int_0^\infty
  \frac{dt}{t^3}\mbox{Tr}_{\mbox{\scriptsize open}}\left( \frac{1}{2}
  \mbox{\bf P}\left(\frac{1 
  +\left(-1\right)^{F}}{2}\right)\left(-1\right)^{\mbox{\scriptsize \bf S}} 
e^{-2\pi t L_0}\right) , \label{an} \\
{\cal M} & = & c \int_0^\infty
  \frac{dt}{t^3}\mbox{Tr}_{\mbox{\scriptsize open}}\left(\frac{\Omega {\cal
  R}}{2}\mbox{\bf P}\left(\frac{1 
  +\left(-1\right)^{F}}{2}\right)\left(-1\right)^{\mbox{\scriptsize \bf S}} 
e^{-2\pi t L_0}\right). \label{ms}
\end{eqnarray}
Here,
\begin{equation}
\mbox{\bf P} = \left(\frac{1+ \Theta_1 + \cdots + \Theta_1^{\left(
        N-1\right)}}{N}\right) 
\left(\frac{1+ \Theta_2 + \cdots + \Theta_2^{\left(
        M-1\right)}}{M}\right)
\end{equation}
is the projector on states invariant under the orbifold group
$\Z_N\times \Z_M$. {\bf S} denotes the space time fermion number.
In order to compute the contribution due to RR exchange in the tree
channel one needs to compute the parts of expressions (\ref{kb}),
(\ref{an}) and (\ref{ms}) which are given in (\ref{closedopen}). (The
space time fermion number insertion has been taken care of by the
minus sign in the second line in (\ref{closedopen}).) The 
transformation back to the tree channel is performed by replacing $t=
\frac{1}{4l}$ for the Klein bottle, $t=\frac{1}{8l}$ for the M\"obius
strip and $t=\frac{1}{2l}$ for the annulus\cite{Gimon:1996rq}. Finally
RR-charge conservation is imposed by demanding the infrared
($l\rightarrow \infty$) limit of the tree channel expression to be finite.

\section{The $\Z_4 \times \Z_2$ $\OR$--Orientifold}
\label{z4z2}

In this section we discuss the $\Z_4 \times \Z_2$ model in great
detail because in this case all possible subtleties show up; therefore
the other models can be treated briefly in the following
sections.

\begin{figure}[ht]
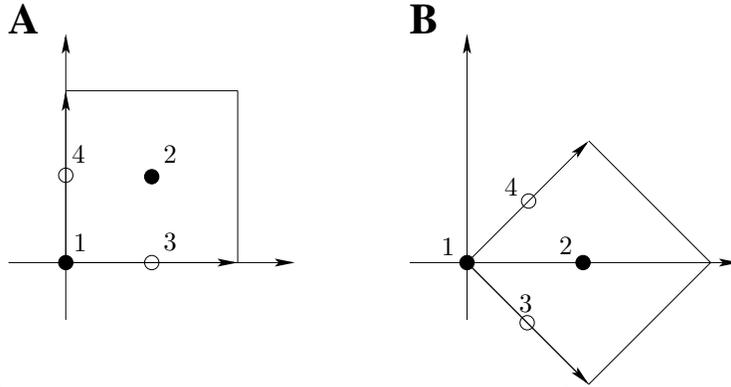
  
\begin{center}
\input fixedpoints42.pstex_t
\end{center}
\vspace{-1cm}
\caption{Lattices for $\Z_4 \times \Z_2$. Black circles denote the
  $\Z_4$ fixed points and white circles the additional $\Z_2$ fixed points.}
\label{fixedpoints42}
\end{figure}

The lattice described by the shifts $\vec{v} = (1/4,-1/4,0)$ for the
$\Z_4$-factor and $\vec{w} = (0,1/2,-1/2)$ for the $\Z_2$-factor of the
orbifold-group is essentially an $SU(2)^6$-lattice, i.e.\ a product of
three tori in the notation of complex compact coordinates. The two
crystallographically allowed
orientations {\bf A} and {\bf B} of one of the three tori with respect to the
reflection ${\mathcal R}$ are shown in
figure~\ref{fixedpoints42}. As will be explained in the following, the
only pertubatively consistent models are given by the choices {\bf
  ABA} and {\bf ABB} for the compact directions\footnote{The choices
  {\bf BAA} and {\bf BAB} are equivalent to these models.}.  

\subsection{The Klein bottle amplitude}
\label{z4z2_kb}

We begin with evaluating the general expression~(\ref{kb}) for the Klein
bottle 1-loop amplitude of the $\Z_4 \times \Z_2$ model by considering
the compact momenta. 
The Kaluza-Klein (KK) and winding (W) states are generally given by
\begin{align}
  \label{kk}
  P &= \frac{\sqrt{2}}{r} \left( m_1 \vec{e_1}^* + m_2 \vec{e_2}^*
  \right) \; , \\
  \label{w}
  W &= \frac{r}{\sqrt{2}\alpha'} \left ( n_1 \vec{e_1} + n_2
  \vec{e_2} \right) \; ,
\end{align}
where $m_i$ and $n_i$ are integers, $\vec{e_i}$ are
the basis vectors of the corresponding torus with radius $r$ and
$\vec{e_i}^*$ are the basis vectors of the dual
torus $(i = 1,2)$. The $SU(2)^2$ lattices in figure~\ref{fixedpoints42}
are spanned by
\begin{align}
  \label{basis_z2_a}
  \vec{e_1}^{\A} &= \left( \begin{array}{c} \sqrt{2} \\ 0
  \end{array} \right) \; , \qquad
  \vec{e_2}^{\A} = \left( \begin{array}{c} 0 \\ \sqrt{2}
  \end{array} \right) \; , \\ 
  \label{basis_z2_b}
  \vec{e_1}^{\B} &= \left( \begin{array}{c} 1 \\ -1
  \end{array} \right) \; , \qquad
  \vec{e_2}^{\B} = \left( \begin{array}{c} 1 \\ 1
  \end{array} \right) \; ,
\end{align}
with the corresponding dual basis. For the Kaluza-Klein and winding states
invariant under $\OR$ one gets
\begin{align}
  \label{kkw_z2_a}
  P^{\A} &= \frac{m}{r} \left( \begin{array}{c} 1 \\ 0
  \end{array} \right) \; , \qquad
  W^{\A} = \frac{nr}{\alpha'} \left( \begin{array}{c} 0 \\ 1
  \end{array} \right) \; , \\
  \label{kkw_z2_b}
  P^{\B} &= \frac{\sqrt{2}m}{r} \left( \begin{array}{c} 1 \\ 0
  \end{array} \right) \; , \qquad
  W^{\B} = \frac{\sqrt{2}nr}{\alpha'} \left( \begin{array}{c} 0 \\ 1
  \end{array} \right) \; , 
\end{align}
where $m,n$ are integers. 
As a consequence of relation~(\ref{durchziehen}) the states~(\ref{kkw_z2_a})
and~(\ref{kkw_z2_b}) are invariant under the insertions $\OR \Theta_1^{k_1}
\Theta_2^{k_2}$ for $k_1 = 0,2$ and $k_2 = arbitrary$; when
{\bf A} and {\bf B} are exchanged, these states 
are invariant under insertions with $k_1 = 1,3$ and $k_2 =
arbitrary$. From
\begin{equation}
  \label{plr}
  p_{L,R} = P \pm W
\end{equation}
for the closed string, it follows that the lattice contribution to the
1-loop amplitude for the Klein bottle is $\LAT[1,1]$ for {\bf A}-states
and $\LAT[2,2]$ for {\bf B}-states, where the notation is taken
from\cite{Blumenhagen:2000md,Blumenhagen:2000ev} and explained in
appendix~\ref{appa}. In general, 
lattice contributions only appear for untwisted tori.

The calculation of the oscillator contributions to~(\ref{kb})
simplifies if one takes into account that the RR-exchange in the
tree level is given by the
trace over the NSNS-sector with the insertion $(-1)^F$ in the
1-loop channel. Furthermore,
the elements of the orbifold group $\Z_4 \times \Z_2$ act as the unit
operator on the oscillator states which contribute to the trace
because $\OR$-invariance leads to a cancellation of the phases given
in equation~(\ref{phases}) between left- and right-movers. This means
that the oscillator contributions are equal for any insertion from
the orbifold group. Although the numerical results may be zero from
case to case, all the twisted sectors formally show up in the amplitude,
because $\OR$ does not exchange them. The last ingredients we need are
the multiplicities $\chi_{\KB}^{(n_1,k_1)(n_2,k_2)}$ of the $\Theta_1^{n_1}
\Theta_2^{n_2}$-twisted fixed points which are invariant under the
insertion $\OR
\Theta_1^{k_1} \Theta_2^{k_2}$. Consider e.g.\ the second torus $T_2$
twisted by $\Theta_2$. In the {\bf A}-lattice, two of the four fixed
points are interchanged under $\OR \Theta_1^{k_1} \Theta_2^{k_2}$ when
$k_1 = 0,2$ and $k_2 = arbitrary$. In the {\bf B}-lattice all four
fixed points are invariant under these insertions, such that 
$\chi_{\KB}^{(0,0)(1,k_2)} = \chi_{\KB}^{(0,2)(1,k_2)} = 2 (4)$ for an {\bf
  A}({\bf B})-type $T_2$. The resulting multiplicities are shown in
table~\ref{multiplicities_kb}.
\renewcommand{\arraystretch}{1.3}
\begin{table}[ht]
  \begin{center}
    \begin{equation*}
      \begin{array}{|l||c|c|} \hline
        \chi_{\KB}           & \mbox{\bf ABA} & \mbox{\bf ABB} \\
        \hline\hline 
        (0,k_1)(0,k_2)       & 1              & 1 \\ \hline
        (2n_1+1,k_1)(0,k_2)  & 4              & 4 \\ \hline
        (2,k_1)(0,k_2)       & 8              & 8 \\ \hline
        (0,2k_1)(1,k_2)      & 8              & 4 \\ \hline
        (0,2k_1+1)(1,k_2)    & 16             & 8 \\ \hline
        (2n_1+1,k_1)(1,k_2)  & 16             & 8 \\ \hline
        (2,2k_1)(1,k_2)      & 16             & 8 \\ \hline
        (2,2k_1+1)(1,k_2)    & 8              & 4 \\ \hline
      \end{array}
    \end{equation*}
  \end{center}
  \caption{Multiplicities of the fixed points for $\Z_4 \times \Z_2$.}
  \label{multiplicities_kb}
\end{table}
Considering all this, we can evaluate the Klein bottle 1-loop
amplitude~(\ref{kb}). The notation is similar
to\cite{Blumenhagen:2000md,Blumenhagen:2000ev} and defined in
appendix~\ref{appa}. We use the fact that for the oscillator
contributions $\KB^{(n_1,k_1)(n_2,k_2)} = \KB^{(n_1,0)(n_2,0)}$ is
valid for all $n_i,k_i$ $(i=1,2)$ as explained above and simplify 
the notation by defining $\KB^{(n_1,n_2)} \equiv
\KB^{(n_1,0)(n_2,0))}$. For the {\bf ABA}-lattice we get
\begin{equation}
  \begin{aligned}
    \label{amplitude_z4z2_kb_loop}
    \KB &= c(1_{\text{RR}} - 1_{\text{NSNS}}) \int_{0}^{\infty} 
    \frac{dt}{t^3} \times \\
    & \Big( \LAT[1,1]^2\LAT[2,2]\KB^{(0,0)} +
      4\LAT[1,1]\KB^{(1,0)} + 8\LAT[1,1]\KB^{(2,0)} +
      4\LAT[1,1]\KB^{(3,0)}  \\
    & + 8\LAT[1,1]\KB^{(0,1)} + 
      16\KB^{(1,1)} + 16\LAT[2,2]\KB^{(2,1)} + 16\KB^{(3,1)} \Big)
    \; .
  \end{aligned}
\end{equation}
For the {\bf ABB}-lattice, one $\LAT[1,1]$ in every term in the second
line of equation~(\ref{amplitude_z4z2_kb_loop}) has to be exchanged for
$\LAT[2,2]$ and the prefactors in the third line have to be divided by
two. 

The modular transformation to the tree-channel $t = \frac{1}{4l}$
yields (see appendix~\ref{appa})  
\begin{equation}
  \begin{aligned}
    \label{amplitude_z4z2_kb_tree}
    \Tilde{\KB} &= 32c(1_{\text{RR}} - 1_{\text{NSNS}}) \int_{0}^{\infty} 
    dl \times \\
    & \Big( \Tilde{\LAT}[4,4]^2\Tilde{\LAT}[2,2]\Tilde{\KB}^{(0,0)} -
      2\Tilde{\LAT}[4,4]\Tilde{\KB}^{(1,0)} -
    4\Tilde{\LAT}[4,4]\Tilde{\KB}^{(2,0)} -
      2\Tilde{\LAT}[4,4]\Tilde{\KB}^{(3,0)}  \\
    & - 4\Tilde{\LAT}[4,4]\Tilde{\KB}^{(0,1)} + 
      4\Tilde{\KB}^{(1,1)} - 4\Tilde{\LAT}[2,2]\Tilde{\KB}^{(2,1)} -
    4\Tilde{\KB}^{(3,1)} \Big) \; .
  \end{aligned}
\end{equation}
For the {\bf ABB}-lattice, one $\Tilde{\LAT}[4,4]$ in every term in the second
line of equation~(\ref{amplitude_z4z2_kb_tree}) again has to be exchanged for
$\Tilde{\LAT}[2,2]$ and equation~(\ref{amplitude_z4z2_kb_tree}) has to be
multiplied by an overall factor of 1/2. We realize that the complete
projector in the sense of\cite{Blumenhagen:2000md} and
section~\ref{setup} shows up, 
because all possible insertions of the orbifold group appear, only
untwisted sectors contribute and the prefactors are given by
\begin{equation}
  \label{complete}
  \prod_{k_1v_i + k_2 w_i \neq 0\, ; \; i=1,2,3} \big
  ( -2\sin(\pi k_1v_i + \pi k_2w_i) \big) \; ,
\end{equation}
as expected. At this point we can clarify, why only the models {\bf
  ABA} and {\bf ABB} (and the equivalent models {\bf BAA} and {\bf
  BAB}) are perturbatively consistent. The lattice contribution of
e.g.\ {\bf AAA} is changed to {\bf BBA} by the insertion of
$\Theta_1$ and there is no way to get the complete projector. From
the same argument it follows that only the orbifold groups $\Z_2
\times \Z_2$, $\Z_4 \times \Z_2$, $\Z_3 \times \Z_3$ and $\Z_6
\times \Z_3$ can lead to perturbatively consistent solutions.

\subsection{The annulus amplitude}
\label{z4z2_an}

\begin{figure}[ht]
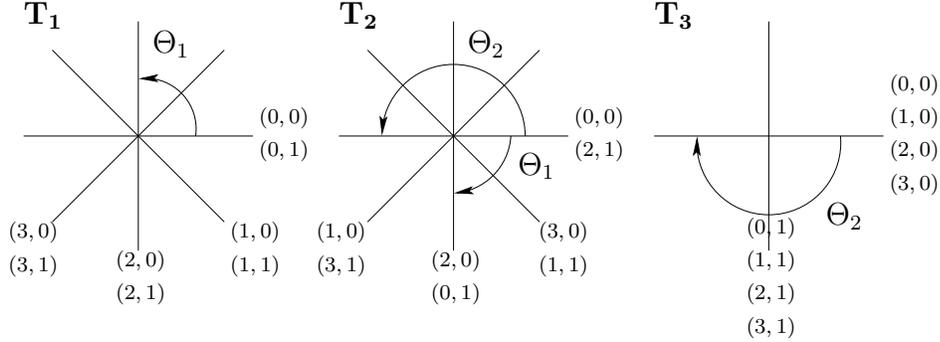
  
\begin{center}
\input brane_conf42.pstex_t
\end{center}
\caption{Arrangement of branes and action of the orbifold group for
  $\Z_4 \times \Z_2$. The branes are labelled by $(i_1,i_2) =
  (0,0), \ldots ,(3,1)$ mod $(4,2)$. 
  This is convenient in the sense that the
  brane $(i_1,i_2)$ is obtained by rotating the compact real axes with
  $\Theta_1^{-i_1/2}\Theta_2^{-i_2/2}$, see also section~\ref{setup}.} 
\label{brane_conf42}
\end{figure}

To cancel the tadpoles which arise in the Klein bottle amplitude we
need to introduce D-branes. As was found
in\cite{Blumenhagen:2000md,Blumenhagen:2000ev} and explained in
section~\ref{setup} for the orientifold models under consideration we
have to introduce D-6-branes rotated by half the 
angles which are given by the elements of the orbifold group. For
$\Z_4 \times \Z_2$ this leads to a configuration of eight different
D-6-branes whose locations in the three compact tori are shown in
figure~\ref{brane_conf42}. For simplicity we restrict ourselves to the
case where the D-branes are located at the fixed points. 
Writing down the mode expansions e.g.\ for
an open string stretching from brane $(0,0)$ to brane
$(1,0)$ (where the brane $(1,0)$ is rotated by $\Theta_1^{-1/2}$
with respect to the brane $(0,0)$) one  
realizes that the modings are the same as for the closed string
twisted by $\Theta_1^{-1}$. Therefore it is convenient to call these
kinds of open strings ``twisted sectors''\cite{Blumenhagen:2000md}, as
we will do in the following. Using these conventions, an open string
stretching from brane $(i_1,i_2)$ to brane $(i_1 - n_1,i_2 - n_2)$
belongs to the $\Theta_1^{n_1}\Theta_2^{n_2}$-twisted sector.

\begin{figure}[ht]
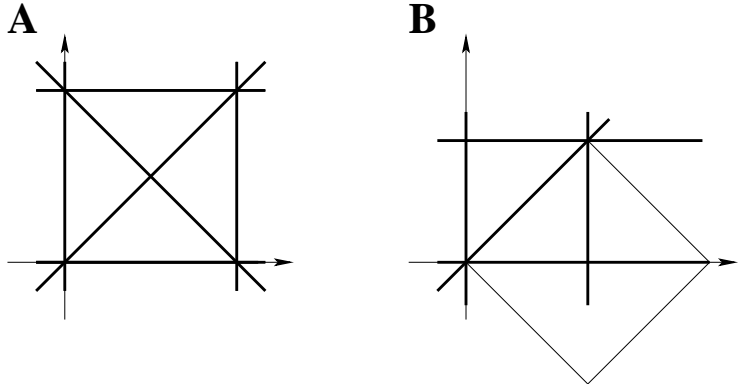
  
\begin{center}
\input intersection42.pstex_t
\end{center}
\vspace{-1cm}
\caption{Location of branes in the fundamental cells for $\Z_4 \times \Z_2$}
\label{intersection42}
\end{figure}

For the open string the compact momenta $p_{open}$ are given by the distance of
parallel D-6-branes in the corresponding directions. Therefore we
consider the location of the branes in the fundamental cells of the
lattices {\bf A} and {\bf B}, see figure~\ref{intersection42}.
Starting from brane $(0,0)$,
we get $p^{\A}_{open} = P^{\A}$ for the directions 4,6,8 and $p^{\A}_{open}
= W^{\A}$ for the directions 5,7,9 as well as $p^{\B}_{open} = \frac{1}{2}
P^{\B}$ (4,6,8) and  $p^{\B}_{open} = \frac{1}{2} W^{\B}$ (5,7,9), see
equations~(\ref{kkw_z2_a}) and~(\ref{kkw_z2_b}). 
It follows, that 
the lattice contribution to the 1-loop amplitude for the annulus is
given by $\LAT[2,2]$ for an {\bf A}-torus and by $\LAT[1,1]$ for a
{\bf B}-torus.
The compact momenta are nonzero again only for untwisted tori, i.e.\
tori where the twist acts trivially. But in
addition, the D-branes have to be invariant under insertions of the
elements of the orbifold group, therefore only the insertions
$\unity$, $\Theta_1^2$, $\Theta_2$ and $\Theta_1^2 \Theta_2$ yield
non-vanishing compact momenta on untwisted tori.

Calculating the oscillator contributions to~(\ref{an}) we again focus
on the RR-exchange in the tree channel which is given by the trace over
the NS-sector with $(-1)^F$ insertion in the 1-loop amplitude. All
twisted sectors appear, each in combination with the insertions
$\unity$, $\Theta_1^2$, $\Theta_2$ and $\Theta_1^2 \Theta_2$ which
leave the D-branes 
invariant, as stated above. In contrast to the Klein bottle amplitude
the phases arising from the insertions are not cancelled. Moreover, the
representation matrices of the orbifold group have to be taken into
account. These matrices are unitary $M \times M$ matrices, where $M$ is
the number of arrangements 
that will be fixed by the tadpole cancellation conditions. In the amplitude
the matrices appear as
\begin{equation}
  \label{cp_z4z2_an}
  \text{tr} \gamma^{(i_1-n_1,i_2-n_2)}_{k_1k_2} \,
   \text{tr} \gamma^{(i_1,i_2)-1}_{k_1k_2} \; ,
\end{equation}
where $(i_1,i_2)$ labels the eight different branes  and
$n_1$ and $n_2$ indicate the 
$\Theta_1^{n_1}\Theta_2^{n_2}$-twisted sector, as explained
above.
$\gamma^{(i_1,i_2)}_{k_1k_2}$ is an abbreviation for 
$\gamma^{(i_1,i_2)}_{\Theta_1^{k_1}\Theta_2^{k_2}}$. 
This leads to a factor of $8M^2$
for the untwisted and twisted sectors without insertions (``without''
means insertion of $\unity$ in the case of the annulus), where the 8
arises from the number of 
branes in the arrangement (see figure~\ref{brane_conf42}). In fact, all the
calculations can be done starting with brane $(0,0)$ and inserting
factors of 8 appropriately, because the other branes lead to the same
amplitudes.

The terms
with insertions lead to twisted sector tadpoles in the tree-channel,
which cannot be cancelled by the other diagrams. This yields the
twisted sector tadpole cancellation conditions
\begin{equation}
  \label{twisted_tp}
  \text{tr} \gamma^{(i_1,i_2)}_{20} = \text{tr} \gamma^{(i_1,i_2)}_{01} =
     \text{tr} \gamma^{(i_1,i_2)}_{21} = 0
\end{equation}
for all $(i_1,i_2)$, similar to\cite{Gimon:1996rq}.  

The analogue to
the multiplicities of fixed points in the closed string are the
intersection numbers of the D-branes in the open string case. The
intersection numbers are given by the number of times that the branes
intersect within the fundamental cell and can be read off easily from
figure~\ref{intersection42}. Starting with brane $(0,0)$, a
$\Z_2$-twisted {\bf B}-type torus 
contributes a factor of two, whereas a $\Z_2$-twisted {\bf A}-type
torus and $\Z_4$-twisted tori of both types contribute a factor of
one\footnote{Remember that we are discussing strings starting on brane
  $(0,0)$.}.
Again, only the points invariant under
insertions contribute, thus in the case of the annulus only sectors
without insertions appear in the amplitude and
it is sufficient to consider the multiplicities
$\chi_{\AN}^{(n_1,n_2)} \equiv \chi_{\AN}^{(n_1,0)(n_2,0)}$ which are
given in table~\ref{multiplicities_an}.
\renewcommand{\arraystretch}{1.3}
\begin{table}[ht]
  \begin{center}
    \begin{equation*}
      \begin{array}{|l||c|c|} \hline
        \chi_{\AN}           & \mbox{\bf ABA} & \mbox{\bf ABB} \\
        \hline\hline 
        (0,0)  & 1             & 1 \\ \hline
        (1,0)  & 1             & 1 \\ \hline
        (2,0)  & 2             & 2 \\ \hline
        (3,0)  & 1             & 1 \\ \hline
        (0,1)  & 2             & 4 \\ \hline
        (1,1)  & 1             & 2 \\ \hline
        (2,1)  & 1             & 2 \\ \hline
        (3,1)  & 1             & 2 \\ \hline
      \end{array}
    \end{equation*}
  \end{center}
  \caption{Intersection numbers in the annulus for $\Z_4 \times \Z_2$.}
  \label{multiplicities_an}
\end{table}
Now we have all the ingredients to write down the annulus 1-loop
  amplitude for the {\bf ABA}-lattice 
\begin{equation}
  \begin{aligned}
    \label{amplitude_z4z2_an_loop}
    \AN &= M^2 \, \frac{c}{4} \, (1_{\text{RR}} - 1_{\text{NSNS}})
    \int_{0}^{\infty}
    \frac{dt}{t^3} \times \\
    & \Big( \LAT[1,1]\LAT[2,2]^2\AN^{(0,0)} +
      \LAT[2,2]\AN^{(1,0)} + 2\LAT[2,2]\AN^{(2,0)} +
      \LAT[2,2]\AN^{(3,0)}  \\
    & + 2\LAT[2,2]\AN^{(0,1)} + 
      \AN^{(1,1)} + \LAT[1,1]\AN^{(2,1)} + \AN^{(3,1)} \Big)
    \; ,
  \end{aligned}
\end{equation}
where we used the simplified notation $\AN^{(n_1,n_2)} \equiv
\AN^{(n_1,0)(n_2,0)}$ (see appendix~\ref{appa}).
All the terms with insertions vanish due to the
twisted tadpole cancellation condition~(\ref{twisted_tp}) and
therefore do not appear in equation~(\ref{amplitude_z4z2_an_loop}). For
the {\bf ABB}-lattice one $\LAT[2,2]$ in every term in the second
line of equation~(\ref{amplitude_z4z2_an_loop}) has to be exchanged for
$\LAT[1,1]$ and the prefactors in the third line have to be multiplied by
two. 

Performing the modular transformation $t = \frac{1}{2l}$ leads to
\begin{equation}
  \begin{aligned}
    \label{amplitude_z4z2_an_tree}
    \Tilde{\AN} &= \frac{c}{8} \, M^2 \, (1_{\text{RR}} - 1_{\text{NSNS}})
    \int_{0}^{\infty}  dl \times \\
    & \Big( \Tilde{\LAT}[1,1]^2\Tilde{\LAT}[2,2]\Tilde{\AN}^{(0,0)} -
      2\Tilde{\LAT}[1,1]\Tilde{\AN}^{(1,0)} -
    4\Tilde{\LAT}[1,1]\Tilde{\AN}^{(2,0)} -
      2\Tilde{\LAT}[1,1]\Tilde{\AN}^{(3,0)}  \\
    & - 4\Tilde{\LAT}[1,1]\Tilde{\AN}^{(0,1)} + 
      4\Tilde{\AN}^{(1,1)} - 4\Tilde{\LAT}[2,2]\Tilde{\AN}^{(2,1)} -
    4\Tilde{\AN}^{(3,1)} \Big) \; .
  \end{aligned}
\end{equation}
For the {\bf ABB}-lattice, one $\Tilde{\LAT}[1,1]$ in every term in the second
line of equation~(\ref{amplitude_z4z2_an_tree}) has to be exchanged for
$\Tilde{\LAT}[2,2]$ and the whole amplitude has to to be multiplied by
two. Again, the complete projector shows up.

\subsection{The M\"obius strip amplitude}
\label{z4z2_ms}

In the case of the M\"obius strip, the compact momenta for a {\bf
  B}-torus have to be doubled in the 5,7,9 directions because of the
$\OR$-projection, therefore one gets $\LAT[2,2]$ for an {\bf A}-torus
and $\LAT[1,4]$ for a {\bf B}-torus. Again, the lattice
contributions only appear for untwisted tori and for insertions which
leave the D-6-branes invariant.

Which strings contribute to the M\"obius strip one-loop amplitude?
Let us denote a $\Theta_1^{n_1}\Theta_2^{n_2}$-twisted string
by $[(i_1,i_2)(i_1-n_1,i_2-n_2)]$ as explained in
section~\ref{z4z2_an} and consider the action of the insertion $\OR
\Theta_1^{k_1}\Theta_2^{k_2}$ thereupon:
\begin{equation}
  \begin{array}{ll}
    \lefteqn{[(i_1,i_2)(i_1-n_1,i_2-n_2)]} & \\
    \rTo^{\Theta_1^{k_1}\Theta_2^{k_2}} 
    & [(i_1+2k_1,i_2+2k_2)(i_1-n_1+2k_1,i_2-n_2+2k_2)] \\
    \rTo^{\;\;\;\;\; \mathcal R \;\;\;\;\;}
    & [(-i_1-2k_1,-i_2-2k_2)(-i_1+n_1-2k_1,-i_2+n_2-2k_2)] \\
    \rTo^{\;\;\;\;\; \Omega \;\;\;\;\;}
    & [(-i_1+n_1-2k_1,-i_2+n_2-2k_2)(-i_1-2k_1,-i_2-2k_2)] \; .
  \end{array}
\end{equation}
Since $n_2, k_2 = 0,1$ (mod 2), the condition $i_2 = -i_2+n_2-2k_2$ (mod
2) is equivalent to $2i_2 = n_2$ (mod 2), thus only
$\Z_2$-untwisted sectors 
(i.e.\ $n_2=0$) with $k_2 = arbitrary$ contribute to the
amplitude. The condition \mbox{$i_1 = -i_1+n_1-2k_1$ (mod 4)} implies e.g.\ for
the brane with $i_1 = 0$ that the sectors $n_1 = 0$ with $k_1 = 0,2$
and $n_1 = 2$ with $k_1 = 1,3$ contribute to the amplitude. To
summarize, for $(i_1,i_2) = (0,0)$ the sectors $(n_1,k_1)(n_2,k_2) =
(0,0)(0,0)$, $(0,0)(0,1)$, $(0,2)(0,0)$,  $(0,2)(0,1)$, $(2,1)(0,0)$,
$(2,1)(0,1)$, $(2,3)(0,0)$ and $(2,3)(0,1)$ contribute. The other branes
of the arrangement in figure~\ref{brane_conf42} get contributions from
different sectors, but the resulting amplitude is the same, 
therefore we can restrict the
calculation to the case $(i_1,i_2) = (0,0)$ and insert factors of 8
appropriately, again. 

Now the representation matrices of the orientifold group have to be
taken into account. For the brane $(i_1,i_2)$ in the sector
$(n_1,k_1)(n_2,k_2)$ they appear as
\begin{equation}
  \label{cp_z4z2_ms}
  \CP{i_1-n_1}{i_2-n_2}{i_1}{i_2}{k_1}{k_2} \; .
\end{equation}
Since only untwisted and $\Theta_1^2$-twisted sectors appear, we
abbreviate $a^{(n_1)}_{k_1k_2}$ $\equiv$ 
$\CP{n_1}{0}{0}{0}{k_1}{k_2}$. The multiplicities $\chi_{\MS}$ can be
obtained in the same way as in the case of the annulus, because all
intersection points are invariant under $\OR$. We get  
$\chi_{\MS} = 2$ in the $\Theta_1^{2}$-twisted
sectors and $\chi_{\MS} = 1$ in the other sectors.

This leads to the M\"obius strip 1-loop amplitude for the {\bf
  ABA}-lattice 
\begin{equation}
  \begin{aligned}
    \label{amplitude_z4z2_ms_loop}
    \MS &= - \frac{c}{4} \, (1_{\text{RR}} - 1_{\text{NSNS}})
    \int_{0}^{\infty}
    \frac{dt}{t^3} \times \\
    & \Big( a^{(0)}_{00} \LAT[1,4]\LAT[2,2]^2\MS^{(0,0)(0,0)}
            + a^{(0)}_{01} \LAT[2,2]\MS^{(0,0)(0,1)} \\
            & + 2a^{(2)}_{10} \LAT[2,2]\MS^{(2,1)(0,0)}
            + 2a^{(2)}_{11} \MS^{(2,1)(0,1)}
            + a^{(0)}_{20} \LAT[2,2]\MS^{(0,2)(0,0)} \\
            & + a^{(0)}_{21} \LAT[1,4]\MS^{(0,2)(0,1)}
            + 2a^{(2)}_{30} \LAT[2,2]\MS^{(2,3)(0,0)}
            + 2a^{(2)}_{31} \MS^{(2,3)(0,1)} \Big) \; .
  \end{aligned}
\end{equation}
For the {\bf ABB}-lattice, one $\LAT[2,2]$ in each term
of equation~(\ref{amplitude_z4z2_ms_loop}) with $a^{(n_1)}_{k_1k_2} =
a^{(n_1)}_{k_10}$ has to be exchanged for $\LAT[1,4]$.

Performing the transformation to the tree-channel $t = \frac{1}{8l}$
(see appendix~\ref{appa}) yields
\begin{equation}
  \begin{aligned}
    \label{amplitude_z4z2_tree}
    \Tilde{\MS} &= - 4c \, (1_{\text{RR}} - 1_{\text{NSNS}})
    \int_{0}^{\infty} dl \times \\
    & \Big( a^{(0)}_{00} \Tilde{\LAT}[8,2]\Tilde{\LAT}[4,4]^2
                                 \Tilde{\MS}^{(0,0)}
            - 2a^{(2)}_{30} \Tilde{\LAT}[4,4]\Tilde{\MS}^{(1,0)}\\
            & + 4a^{(0)}_{20} \Tilde{\LAT}[4,4]\Tilde{\MS}^{(2,0)}
            -2a^{(2)}_{10} \Tilde{\LAT}[4,4]\Tilde{\MS}^{(3,0)}
            + 4a^{(0)}_{01} \Tilde{\LAT}[4,4]\Tilde{\MS}^{(0,1)} \\ 
            & + 4a^{(2)}_{31} \Tilde{\MS}^{(1,1)}
            + 4a^{(0)}_{21} \Tilde{\LAT}[8,2]\Tilde{\MS}^{(2,1)} 
            + 4a^{(2)}_{11} \Tilde{\MS}^{(3,1)}  \Big) \; .
  \end{aligned}
\end{equation}
For the {\bf ABB}-lattice, one $\LAT[4,4]$ in each term
of equation~(\ref{amplitude_z4z2_tree}) with $a^{(n_1)}_{k_1k_2} =
a^{(n_1)}_{k_10}$ has to be exchanged for $\LAT[8,2]$.

To obtain the complete projector and to cancel the untwisted tadpoles
from the other diagrams, the $\gamma$-matrices have to fulfill the
conditions
\begin{equation}
  \label{cp_tp}
    a^{(0)}_{00} = a^{(2)}_{10} = -a^{(0)}_{20}  =  a^{(2)}_{30} =
    -a^{(0)}_{01} = -a^{(2)}_{11} = -a^{(0)}_{21} = a^{(2)}_{31} = M \; 
\end{equation}
and the untwisted tadpole cancellation condition reads
\begin{equation}
  \label{untwisted_tp}
  \begin{aligned}
    \text{{\bf ABA}:} & \qquad \left[ M - 16 \right]^2 = 0 \; ,\\
    \text{{\bf ABB}:} & \qquad \left[ M - 8 \right]^2 = 0 \; ,
  \end{aligned}
\end{equation}
which fixes the number of arrangements shown in
figure~\ref{brane_conf42} to be 
16(8) for the {\bf ABA} ({\bf ABB}) lattice, respectively. The
conditions~(\ref{cp_tp}) are valid for the brane $(i_1,i_2) =
(0,0)$. For some other brane $(i_1,i_2)$ one has to replace the insertions
$\Theta_1^{k_1} \Theta_2^{k_2}$ in~(\ref{cp_tp}) by
$\Theta_1^{k_1+i_1} \Theta_2^{k_2+i_2}$.

\subsection{The closed string spectrum}
\label{z4z2_closed}

The massless spectrum is found by symmetrizing the massless states
which satisfy the GSO-projection conditions~(\ref{gso}) with respect
to $\OR$, $\Theta_1$ and $\Theta_2$. $\Omega$ exchanges left- and
right-movers and is defined following the convention
of\cite{Gimon:1996rq}
\begin{equation}
  \Omega \alpha_{r} \Omega^{-1} = \Tilde{\alpha}_{r} \; , \qquad
  \Omega \psi_{r} \Omega^{-1} = \Tilde{\psi}_{r}\; , \qquad
  \Omega \Tilde{\psi}_{r} \Omega^{-1} = -\psi_{r}
\end{equation}
for integer and half-integer $r$. For the action of ${\mathcal R}$,
$\Theta_1$ and $\Theta_2$ see section~\ref{setup}.
In the following we denote the NSNS vacuum by $|0\rangle$ and e.g.\ the
R state $|\frac{1}{2},\frac{1}{2},\frac{1}{2},\frac{1}{2}\rangle_L$ by
$|++++\rangle_L$. The states are given up to normalization.
In the untwisted sector we find the massless states
\begin{equation*}
  \begin{array}{lll}
    \text{NSNS:}
    & (\psi^{\mu} \Tilde{\psi}^{\nu} + \psi^{\nu}
    \Tilde{\psi}^{\mu}) |0\rangle
    & \text{graviton $+$ dilaton (m.i.)} \\
    \ & \psi^i \Tilde{\psi}^{\Bar{i}} |0\rangle \; , \;
    \psi^{\Bar{i}} \Tilde{\psi}^i |0\rangle 
    & \text{$(i = 1,2,3; \; \Bar{i} = \Bar{1},\Bar{2},\Bar{3})$ 6
      scalars (m.i.)} \\  
    \ & (\psi^3\Tilde{\psi}^3 + \psi^{\Bar{3}}
    \Tilde{\psi}^{\Bar{3}}) |0\rangle
    & \text{1 scalar} \\
    \text{RR:}
    & {\scriptstyle |++++\rangle_L |-+++\rangle_R \; - \;
      |----\rangle_L |+---\rangle_R}
    & \text{axion (m.i.)} \\ 
    \ & {\scriptstyle |-++-\rangle_L |+++-\rangle_R \; - \; 
      |+--+\rangle_L |---+\rangle_R}
    & \text{1 scalar}
  \end{array}
\end{equation*}
where (m.i.) stands for ``model independent'' states, i.e.\ states
which are present independent of the orbifold group.
To summarize, the untwisted massless closed string spectrum contains the
${\cal N}=1$ supergravity multiplet in $D=4$ and 4C, where C denotes the
chiral multiplet.

In the $\Theta_1^{n_1}\Theta_2^{n_2}$-twisted sectors the masses are given by
\begin{equation}
  \frac{\alpha'}{4} m_{L,R}^2 = N_{L,R} + \frac{1}{2} q_{L,R}^2 +
  E_{vac} - \frac{1}{2} \; ,
\end{equation}
with
\begin{equation}
  q_{L,R} = \left\{ \begin{array}{ll} 
                      \left(0,\pm (n_1 \vec{v} + n_2 \vec{w}) \right) 
                           \qquad & \text{(NS)} \\
                      \left(\frac{1}{2},\frac{1}{2} \pm (n_1 \vec{v} + n_2
                        \vec{w}) \right) \qquad & \text{(R)}
                    \end{array} \right.
\end{equation}
and
\begin{equation}
  E_{vac} = \frac{1}{2} \sum_{i=1,2,3} |n_1v_i + n_2w_i|(1 - |n_1v_i +
  n_2w_i|) \; ,
\end{equation}
where one has to take care of $0 \leq |n_1v_i + n_2w_i| < 1$. We
discuss the $\Theta_1^2$-twisted sector explicitly, the other sectors
are obtained in a similar manner. 

In the $\Theta_1^2$-twisted NSNS sector, the massless states which
fulfill the GSO-projection~(\ref{gso}) are found to be
\begin{equation*}
\begin{array}{ll}
  \textstyle
  |0,\frac{1}{2},\frac{1}{2},0\rangle_L \equiv
  |1\rangle_{NS} \; , &
  |0,-\frac{1}{2},-\frac{1}{2},0\rangle_L \equiv
  |2\rangle_{NS} \; , \\
  |0,-\frac{1}{2},-\frac{1}{2},0\rangle_R \equiv
  |\Tilde{1}\rangle_{NS} \; ,  &
  |0,\frac{1}{2},\frac{1}{2},0\rangle_R \equiv
  |\Tilde{2}\rangle_{NS} \; , 
\end{array}
\end{equation*}
such that we get the two  massless ground states 
$|1\Tilde{1}\rangle_{NSNS}$ and $|2\Tilde{2}\rangle_{NSNS}$. The other
two possible combinations are not invariant under $\Theta_2$. Furthermore
we have to consider the action of $\OR$, $\Theta_1$ and $\Theta_2$ on
the fixed points. Since the torus $T_3$ is untwisted, the discussion is valid
for both the {\bf ABA} and the {\bf ABB} lattice. For the fixed point
structure see figure~\ref{fixedpoints42}. In the tori $T_1$ and $T_2$, the
fixed points 3 and 4 are interchanged by $\Theta_1$. In addition, the
fixed points 3 and 4 in the torus $T_2$ are interchanged by
$\OR$. Under the action of $\Theta_2$ all the fixed points are
invariant. In the following e.g.\ $\{13\}$ denotes the fixed point built
from fixed points 1 of $T_1$ and 3 of $T_2$. The fixed points \{11\},
\{12\}, \{21\} and \{22\} are invariant under $\OR$ and
$\Theta_{1}$. The fixed points \{31\}, \{41\} as well as \{32\}, \{42\}
form pairs under $\Theta_{1}$. The fixed points \{13\}, \{14\} as well
as \{23\}, \{24\} form pairs 
under $\Theta_{1}$ and $\OR$.
The remaining fixed points \{33\}, \{34\}, \{43\} and \{44\}
form a quartet under $\Theta_{1}$ and $\OR$. Symmetrization in the
NSNS sector leads to two scalars for each fixed point, each pair and
the quartet, i.e.\ 18 scalars altogether.

The discussion of the $\Theta_1^2$-twisted RR sector is similar, but
here the action of $\OR$ gives an additional minus sign. Therefore we
have to antisymmetrize between left and right movers, such that we get
no states from the fixed points and pairs mentioned above. The quartet
contributes one vector (V).

Adding the superpartners from the NSR sector, we find
9C + 1V in the $\Theta_1^2$-twisted sector.
The remaining twisted sectors can be treated in a
similar fashion and we obtain the massless twisted closed string
spectrum
\begin{equation}
  \begin{aligned}
    \text{{\bf ABA}:} & \qquad \text{57C + 1V} \; ,\\
    \text{{\bf ABB}:} & \qquad \text{47C + 11V} \;. 
  \end{aligned}
\end{equation}

\subsection{The open string spectrum}
\label{z4z2_open}

In order to determine the open string spectrum we have to count the degrees of
freedom of the Chan-Paton factors for the massless states. Therefore,
we have to find a representation of the orientifold group
which satisfies the tadpole cancellation conditions~(\ref{twisted_tp})
and~(\ref{cp_tp}).
In general, we have to consider the action of the orientifold group on
massless states of the form $|\psi,ij\rangle \lambda^{(a,b)}_{ji}$, where
$\psi$ represents the vacuum together with some combination of
oscillators, $i,j = 1, \ldots, M$ ($M$ is the number of arrangements
of branes) and $\lambda^{(a,b)}$ is the
Chan-Paton matrix for a string starting on brane $a$ and ending on
brane $b$, with $a,b = 1, \ldots, 8$, i.e.\footnote{For notational simplicity
  in the following we label the eight branes in the arrangement shown in
  figure~\ref{brane_conf42} with single numbers as given in
  table~\ref{rep-matrices_OR_z4z2} of appendix~\ref{projective-rep_z4z2}.} 
\begin{equation}
\label{projection}
  (\OR\Theta_1^{k_1}\Theta_2^{k_2}):
  |\psi,ij\rangle\lambda^{(a,b)}_{ji} \longrightarrow 
  |\OR\Theta_1^{k_1}\Theta_2^{k_2} \cdot \psi,ij\rangle 
  \left( \gamma^{(b)}_{\OR k_1k_2}\lambda \gamma^{(a)-1}_{\OR
    k_1k_2} \right)^T_{ji} \; .
\end{equation}
To check whether the twisted tadpole cancellation
conditions~(\ref{twisted_tp}) are satisfied, we need
the representation matrices of the orbifold group which can be obtained
via e.g.
\begin{equation}\label{determine_gamma_Theta}
  (\OR\Theta_1^{k_1-1}\Theta_2^{k_2})(\OR\Theta_1^{k_1}\Theta_2^{k_2}):
  |\psi,ij\rangle\lambda^{(a,b)}_{ji} \; =\;  
  \Theta_1 : |\psi,ij\rangle\lambda^{(a,b)}_{ji}
\end{equation}
which implies 
$\gamma^{(a)}_1 \simeq  \gamma^{(b)-T}_{\OR k_1-1,k_2}
\gamma^{(a)}_{\OR k_1k_2} \;$, 
where ``$\simeq$'' means equal up to an
irrelevant phase. Taking into account all these constraints, we find the
$\gamma$-matrices listed in
appendix~\ref{projective-rep_z4z2}. They form a projective
representation of the orientifold group~\footnote{For a summary on
  projective representations and further references see the appendix
  of~\cite{Klein:2000tf}.}, where  
the $\Z_2 \times \Z_2$ substructure is similar to the model discussed
in\cite{Berkooz:1997dw}, which in turn is T-dual to the $\Z_2 \times \Z_2$
$\OR$-orientifold discussed in section~\ref{zweikreuzzwei}.

In the untwisted NS sector (i.e.\ strings which start and end on the
same brane) the massless states are given by $\psi^m_{-1/2}
|0,ij\rangle \lambda^{(a,a)}_{ji}$ ($m = 0, \ldots , 9)$. 
The string $(1,1)$ is invariant
under $\OR$, $\OR\Theta_1^2$, $\OR\Theta_2$, and
$\OR\Theta_1^2\Theta_2$, where the last symmetry is not independent 
of the first three. Applying equation~(\ref{projection}), we get the
following constraints on the Chan-Paton matrix $\lambda^{(1,1)}$ in
the noncompact directions $\mu = 0, \ldots, 3$
\begin{equation}
  \lambda^{(1,1)} = - \left(\gamma^{(1)}_{\OR00} \lambda^{(1,1)}
                       \gamma^{(1)-1}_{\OR00} \right)^T
                  = - \left(\gamma^{(1)}_{\OR20} \lambda^{(1,1)}
                       \gamma^{(1)-1}_{\OR20} \right)^T
                  = - \left(\gamma^{(1)}_{\OR01} \lambda^{(1,1)}
                       \gamma^{(1)-1}_{\OR01} \right)^T \; ,
\end{equation}
where the minus signs arise from the action of $\OR$ on the massless state.
Using the $\gamma$-matrices listed in appendix~\ref{projective-rep_z4z2} this leads to
\begin{equation}
  \label{projection_11}
  \lambda^{(1,1)} = - \lambda^{(1,1)T}
                  = M_1 \lambda^{(1,1)T} M_1
                  = M_2 \lambda^{(1,1)T} M_2 \; .
\end{equation}
Counting the remaining degrees of freedom of $\lambda^{(1,1)}$ we find
that $\psi^{\mu}_{-1/2} |0,ij\rangle \lambda^{(1,1)}_{ji}$ is a vector
in the adjoint representation of the gauge group $Sp(\frac{M}{4})$. In
a similar manner the compact directions $\psi^{i,\Bar{i}}_{-1/2}
|0,kl\rangle \lambda^{(1,1)}_{lk}$, 
$(i = 1,2,3; \; \Bar{i} = \Bar{1},\Bar{2},\Bar{3})$
yield 3C in the antisymmetric representation of
$Sp(\frac{M}{4})$. 

For the $(2,2)$-string, the symmetries are
$\OR\Theta_1$, $\OR\Theta_1^3$ and $\OR\Theta_1\Theta_2$. Inserting
the corresponding $\gamma$-matrices also leads to
equation~(\ref{projection_11}) for $\lambda^{(2,2)}$,
i.e.\ the same result as for the $(1,1)$-string, but now the gauge
group $Sp(\frac{M}{4})$ constitutes a second factor of a product gauge
group, since the strings are not mapped onto each other by any
symmetry of the theory.

The invariances of the $(3,3)$-string are the same as for the
$(1,1)$-string and lead to the same degrees of freedom. But since the
$(1,1)$-string is mapped onto the $(3,3)$-string by a symmetry of the
theory, namely $\Theta_1$, the $(3,3)$-string is charged under the
same factor of the gauge group and does not contribute any further
matter. Furthermore, we have to check that the additional identities 
$\lambda^{(3,3)} = \alpha \gamma^{(1)}_1 \lambda^{(1,1)}
\gamma^{(1)-1}_1$ ($\alpha = 1$ in the directions $0, \ldots, 3$,
$\alpha = i$ on $T_1$, $\alpha = -i$ on $T_2$ and $\alpha = 1$ on
$T_3$) are consistent with the degrees of freedom found so 
far, which is the case, indeed.

Proceeding in a similar manner for the strings $(4,4), \ldots (8,8)$
we find that the untwisted open string spectrum contains 1V in the
adjoint of the gauge group $\left[ Sp(\frac{M}{4}) \right]^4$ and 3C
in the $[(A,1,1,1) \oplus (1,A,1,1) \oplus (1,1,A,1) \oplus (1,1,1,A)]$,
where $A$ denotes the antisymmetric representation of
$Sp(\frac{M}{4})$ and the four factors of the gauge group arise from
the strings $(1,1)$, $(2,2)$, $(5,5)$ and $(6,6)$, respectively.  
The strings $(3,3)$, $(4,4)$, $(7,7)$ and $(8,8)$ are related to these
strings by the symmetry $\Theta_1$ and do not contribute any further
degrees of freedom, as explained above.

We begin the discussion of the open string twisted sectors with the
$(1,2)$-string. In the sense of the explanation in the beginning of
section~\ref{z4z2_an}, this string forms the $\Theta_1^3$-twisted
sector and the massless states are given by
$\psi^1_{-1/4} |0_{1^3},ij\rangle \lambda^{(1,2)}_{ji}$ and
$\psi^{\Bar{2}}_{-1/4} |0_{1^3},ij\rangle \lambda^{(1,2)}_{ji}$, where
$0_{1^3}$ denotes the $\Theta_1^3$-twisted NS vacuum. The
$(1,2)$-string is invariant under $\Theta_1^2$ and
$\Theta_2$. This leads to the constraints on the Chan-Paton factors 
\begin{equation}
  \lambda^{(1,2)} = \pm \gamma^{(2)}_{20} \lambda^{(1,2)}
                    \gamma^{(1)-1}_{20}
                  = \pm \gamma^{(2)}_{01} \lambda^{(1,2)}
                    \gamma^{(1)-1}_{01} \; ,
\end{equation}
where the signs are unphysical, since we only know that $\Theta_1^4$
and $\Theta_2^2$ act trivially on the $(1,2)$-string
and inserting the corresponding $\gamma$-matrices yields
\begin{equation}
  \lambda^{(1,2)} = \pm M_1 \lambda^{(1,2)} M_1 
                  = \pm M_2 \lambda^{(1,2)} M_2 \;
\end{equation}
where the signs are unphysical and thus arbitrary. Calculating the
degrees of freedom we obtain that the $(1,2)$-string transforms in the
bifundamental $(F,F,1,1)$ of the gauge group. The strings $(2,3)$,
$(3,4)$ and $(4,1)$ are related to the $(1,2)$-string by the
symmetries $\OR$ and $\Theta_1$. Since the massless state is twofold
degenerated, these strings form 2 scalars in the $(F,F,1,1)$. Analogous the
strings $(5,6)$, $(6,7)$, $(7,8)$ and $(8,5)$ form 2 scalars in the
$(1,1,F,F)$. The $\Theta_1$-twisted sector is equivalent to the
$\Theta_1^3$-twisted sector, which contains the strings $(1,4)$,
$(2,1)$ etc. Therefore these sectors together yield 2C
in the $(F,F,1,1) \oplus (1,1,F,F)$. Since only $\Z_4$-twisted
intersection points appear, the multiplicity is one.

The $(1,3)$-string, i.e.\ the $\Theta_1^2$-twisted sector, possesses
the additional symmetries $\OR\Theta_1$ and $\OR\Theta_1^3$, which
lead to the constraints
\begin{equation}
  \lambda^{(1,3)} = - \left( \gamma^{(3)}_{\OR10} \lambda^{(1,3)}
                       \gamma^{(1)-1}_{\OR10} \right)^T
                  = - \left( \gamma^{(3)}_{\OR30} \lambda^{(1,3)}
                       \gamma^{(1)-1}_{\OR30} \right)^T \; ,
\end{equation} 
where the minus signs again arise from the action of $\OR$ on the
massless states. Inserting the corresponding $\gamma$-matrices yields
\begin{equation}
  \lambda^{(1,3)} = - A^T \lambda^{(1,3)T} A = B^T \lambda^{(1,3)T} B
  \; .
\end{equation} 
Counting the degrees of freedom, we find that the $(1,3)$-string
transforms in the $(A,1,1,1)$, where $A$ denotes the antisymmetric
representation of $Sp(\frac{M}{4})$. The $(1,3)$-string is related to
the $(3,1)$-string by $\Theta_1$, just as $(2,4)$ to $(4,2)$, $(5,7)$
to $(7,5)$ and $(6,8)$ to $(8,6)$. The massless states are twofold
degenerated and since $T_1$ and $T_2$ are $\Z_2$-twisted, the
multiplicity is 2 in both the {\bf ABA}- and the {\bf
  ABB}-lattice. Thus the $\Theta_1^2$-twisted sector contributes 2C in
the $[(A,1,1,1) \oplus (1,A,1,1) \oplus (1,1,A,1) \oplus (1,1,1,A)]$ for
both lattices.

The $(1,5)$-string, i.e.\ the $\Theta_2$-twisted sector, is invariant
under $\Theta_1^2$ and $\Theta_2$. Again imposing that the square of
these symmetries act trivially on the massless states, we get the
constraints
\begin{equation}
  \lambda^{(1,5)} = \pm i N_2 \lambda^{(1,5)} M_1 = \pm
  M_1\lambda^{(1,3)T} M_2 \; .
\end{equation} 
This implies that the strings $(1,5)$, $(5,1)$, $(3,7)$ and $(7,3)$,
which are related by $\OR$ and $\Theta_1$, transform in the
$(F,1,F,1)$. Analogous the strings $(2,6)$, $(6,2)$, $(4,8)$ and
$(8,4)$ transform in the $(1,F,1,F)$. The massless states are twofold
degenerated. Now the tori $T_2$ and $T_3$ are $\Z_2$-twisted, thus the
multiplicity is 2 in the {\bf ABA}-lattice and 4 in the {\bf
  ABB}-lattice. Altogether the $\Theta_2$-twisted sector contributes
2C (4C) in the $(F,1,F,1) \oplus (1,F,1,F)$ for the {\bf ABA} ({\bf
  ABB}) lattice.

Proceeding in a similar manner for the remaining twisted sectors we
obtain the open string twisted massless spectrum shown in
table~\ref{z4z2_open_table}. The only pecularity is the fact that
the $\Theta_1 \Theta_2$-twisted sector (and the sector twisted by
$\Theta_1^3 \Theta_2$) has only one massless ground state.
\renewcommand{\arraystretch}{1.3}
\begin{table}[ht]
  \begin{center}
    \begin{equation*}
      \hspace{-0.8cm}
      \begin{array}{|c|c|c|c|} \hline
        \text{twist-sector} & \text{\bf ABA} & \text{\bf ABB}
        & \text{gauge group / matter}  \\ \hline \hline
        \begin{array}{c} \text{untwisted} \\ \ \\ \ \end{array}
        & \multicolumn{2}{|c|}{
          \begin{array}{c} 1\text{V} \\ 3\text{C} \\ \ \end{array}}
        & \begin{array}{c} [Sp(\frac{M}{4})]^4 \\
                           (A,1,1,1) \oplus (1,A,1,1) \\ 
                           \oplus (1,1,A,1) \oplus (1,1,1,A)
                           \end{array} \\ \hline 
        \Theta_1 + \Theta_1^3 & \multicolumn{2}{|c|}{ 2\text{C} }
        & (F,F,1,1) \oplus (1,1,F,F) \\ \hline
        \begin{array}{c} \Theta_1^2 \\ \ \end{array}
        & \multicolumn{2}{|c|}{ \begin{array}{c} 2\text{C}  \\ \ 
        \end{array} } 
        & \begin{array}{c} (A,1,1,1) \oplus (1,A,1,1) \\
                           \oplus (1,1,A,1) \oplus
                           (1,1,1,A) \end{array} \\ \hline
        \Theta_2 & 2\text{C} & 4\text{C}
        & (F,1,F,1) \oplus (1,F,1,F) \\ \hline
        \Theta_1 \Theta_2 + \Theta_1^3 \Theta_2 & 1\text{C} & 2\text{C}
        & (F,1,1,F) \oplus (1,F,F,1) \\ \hline
        \Theta_1^2 \Theta_2 & 1\text{C} & 2\text{C}
        & (F,1,F,1) \oplus (1,F,1,F) \\ \hline
      \end{array}
    \end{equation*}
  \end{center}
  \vspace{-0.5cm}
  \caption{Open string massless spectrum of $\Z_4 \times \Z_2$.}
  \label{z4z2_open_table}
\end{table}
\renewcommand{\arraystretch}{1.0}

\section{The $\Z_2 \times \Z_2$ $\OR$--Orientifold}\label{zweikreuzzwei}
\label{z2z2}

Since the orbifold group $\Z_2 \times \Z_2$ is contained as
a substructure in the $\Z_4 \times \Z_2$ model discussed above, 
the calculation is similar and we do
not have to go into the details again.

\begin{figure}[ht]
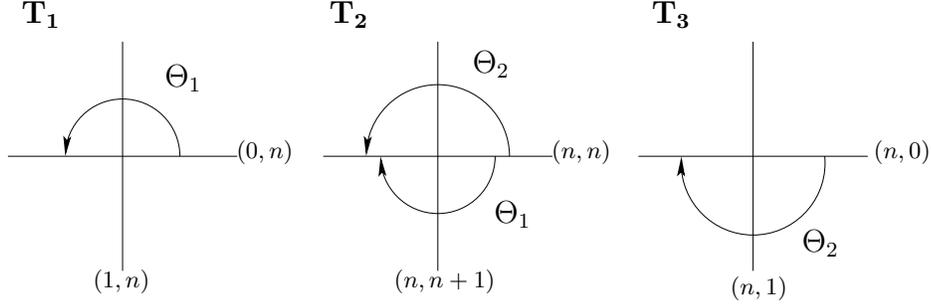
  
\begin{center}
\input brane_conf22.pstex_t
\end{center}
\vspace{-0.8cm}
\caption{Arrangement of branes and action of the orbifold group for
  $\Z_2 \times \Z_2$. Inserting $n=0,1$ yields the four branes
  $(i_1,i_2) = (0,0), (1,0), (0,1), (1,1)$ mod $(2,2)$.} 
\label{brane_conf22}
\end{figure}

The lattice is described by the shifts $\vec{v} = (1/2,-1/2,0)$ for the
first $\Z_2$-factor and $\vec{w} = (0,1/2,-1/2)$ for the second one
and shown in figure~\ref{fixedpoints42}.
The {\bf A}- and {\bf B}-lattice are not
interchanged by any insertion from the orientifold group, thus we obtain
perturbatively consistent and inequivalent solutions for the lattices
{\bf AAA}, {\bf AAB}, {\bf ABB} and {\bf BBB}. The arrangement of
rotated D-6-branes we have to introduce is shown in
figure~\ref{brane_conf22}. For the location of the branes in the
fundamental cells, consider figure~\ref{intersection42} and discard
the diagonal branes. 

Performing the calculation of the amplitudes, we obtain the
untwisted sector tadpole cancellation conditions
\begin{equation}
  \label{untwisted_tp_22}
  \begin{aligned}
    \text{{\bf AAA}:} & \qquad \left[ M - 32 \right]^2 = 0 \; ,\\
    \text{{\bf AAB}:} & \qquad \left[ M - 16 \right]^2 = 0 \; ,\\
    \text{{\bf ABB}:} & \qquad \left[ M - 8 \right]^2 = 0 \; ,\\
    \text{{\bf BBB}:} & \qquad \left[ M - 4 \right]^2 = 0 \; ,
  \end{aligned}
\end{equation}
which fix the number $M$ of arrangements of branes shown in
figure~\ref{brane_conf22}. The remaining tadpole cancellation
conditions yield the representation matrices of the orientifold
group, which can be read off from the $\Z_2 \times \Z_2$ substructure
of tables~(\ref{rep-matrices_OR_z4z2}) and~(\ref{rep-matrices_z4z2})
in appendix~\ref{projective-rep_z4z2}. Using 
these matrices, we again solve the constraints arising from the
symmetries of the various strings and obtain the massless open string
spectrum shown in table~(\ref{z2z2_open}) in appendix~\ref{spectra}. The
massless closed string 
spectrum is shown in table~(\ref{z2z2_closed}) 
in appendix~\ref{spectra}.

Considering the spectrum we can explicitly verify that for the {\bf
  AAA} lattice the $\Z_2
\times \Z_2$ $\OR$-orientifold is T-dual to the model discussed
in\cite{Berkooz:1997dw}. Applying T-duality in the directions of the imaginary
axes of the compact dimensions transforms $\OR$ to $\Omega$ and the
four D-6-branes in figure~\ref{brane_conf22} into one D-9-brane and three
types of D-5-branes.
The {\bf AAB},{\bf ABB} and {\bf BBB} models are T-dual to the $\Z_2
  \times \Z_2$ orientifolds with discrete $B$ field of rank 2,4 and 6,
  respectively, listed in~\cite{Kakushadze:1998eg}. To our knowledge,
  orientifolds with discrete $B$ field 
  have been discussed first in~\cite{Bianchi:1992eu}. For $max(N,M)>2$
  the T-duals are asymmetric orientifolds with nonzero $B$
  field\cite{Blumenhagen:2000fp}. Interestingly, for heterotic 
  theories, which are believed to be connected to open string theories,
  similar observations have been made in\cite{Lauer:1989ax,Lauer:1991tm}.

\section{The $\Z_6 \times \Z_3$ $\OR$--Orientifold}
\label{z6z3}

\begin{figure}[ht]
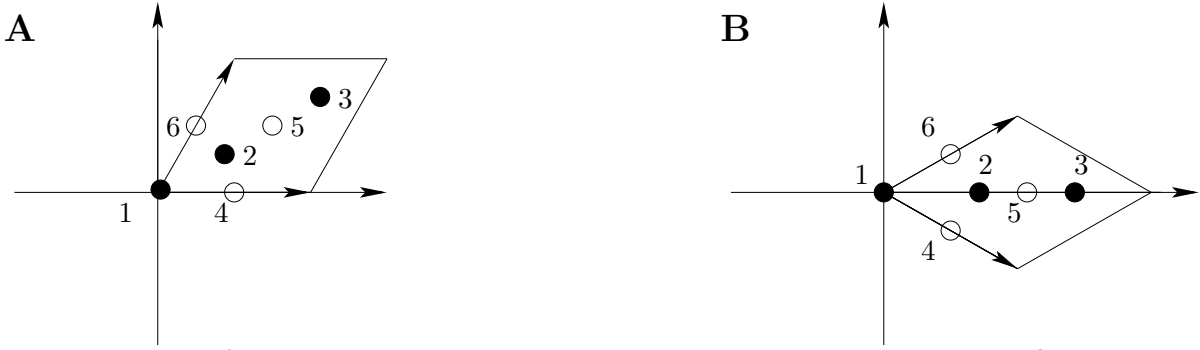
  
\begin{center}
\input fixedpoints63.pstex_t
\end{center}
\vspace{-1cm}
\caption{Lattices for $\Z_6 \times \Z_3$. The black circle in the
  origin denotes the $\Z_6$ fixed point, the other black (white) circles
  denote the additional $\Z_3$ ($\Z_2$) fixed points.}
\label{fixedpoints63}
\end{figure}

\begin{figure}[ht]
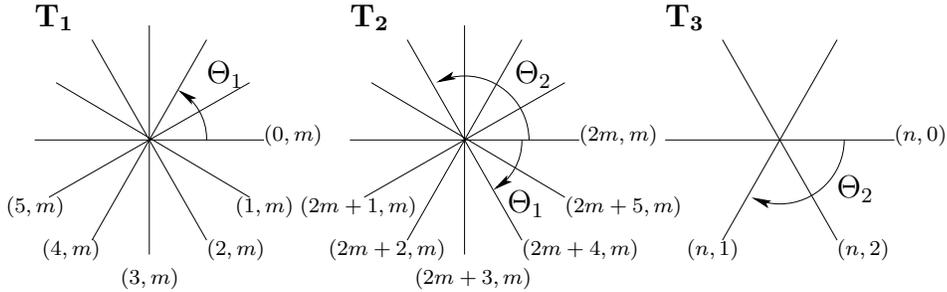
  
\begin{center}
\input brane_conf63.pstex_t
\end{center}
\vspace{-0.8cm}
\caption{Arrangement of branes and action of the orbifold group for
  $\Z_6 \times \Z_3$.  Inserting $n=0, \ldots, 5$ and $m = 0,1,2$ yields
  18 branes.}  
\label{brane_conf63}
\end{figure}

The lattice for this model is generated by the shifts $\vec{v} =
(1/6,-1/6,0)$ and $\vec{w} = (0,1/3,-1/3)$ and shown in
figure~\ref{fixedpoints63}. The {\bf A}- and {\bf B}-lattice are
interchanged by $\OR \Theta_1$, thus we get solutions for the
lattices {\bf ABA} and {\bf ABB}.
To cancel the tadpoles from the closed string we have to introduce the
arrangement of 18 rotated branes shown in figure~\ref{brane_conf63}
and the number $M$ of arrangements is fixed by the untwisted tapole
cancellation condition
\begin{equation}
  \label{untwisted_tp_63}
    \left[ M - 4 \right]^2 = 0 \; ,
\end{equation}
which we obtain for both the lattices {\bf ABA} and {\bf ABB}.
Since this model contains less $\Z_2$ substructure than the two models
discussed above, we expect the projective representation of the orientifold
group to be less complicated. In fact it turns out that we get by with
the $\gamma$-matrices of\cite{Gimon:1996rq}, taking into account the
relative signs arising from the tadpole cancellation conditions.
Calculating the massless
open string spectrum shown in table~(\ref{z6z3_open}) in
appendix~\ref{spectra} we have to take
care of the fact that some of the intersection points are not
invariant under various insertions. The massless closed string
spectrum is shown in table~(\ref{z6z3_closed}) in appendix~\ref{spectra}.

\section{The $\Z_3 \times \Z_3$ $\OR$--Orientifold}
\label{z3z3}

The orbifold shifts are now given by $\vec{v} =
(1/3,-1/3,0)$ and $\vec{w} = (0,1/3,-1/3)$. The lattice is the same as
shown in figure~\ref{fixedpoints63}, discarding the
additional $\Z_2$ fixed points. The choices {\bf AAA}, {\bf AAB}, {\bf
  ABB} and {\bf BBB} lead to perturbatively consistent solutions. All
these four models yield the untwisted tadpole cancellation condition
displayed in equation~(\ref{untwisted_tp_63}). The calculation of the
massless open string spectrum is particularly simple in this case,
because the orbifold group contains no $\Z_2$-substructure. Therefore we
can choose the $\gamma$-matrices to be the identity-matrix (up to a
phase), while the relative signs are again fixed by the tadpole
cancellation conditions. Altogether we obtain the massless spectrum
shown in the 
tables~(\ref{z3z3_closed}) and~(\ref{z3z3_open}) in appendix~\ref{spectra}.

\section{Conclusions}

In this paper we presented a class of orientifolds of type IIA
string theory with orbifold group $\Z_N\times \Z_M$. In addition,
symmetry under worldsheet parity $\Omega$ combined with the reflection
${\cal R}$ of three 
directions was imposed. Further, we considered only cases leading to
unbroken ${\cal N}=1$ supersymmetry in the four non compact
directions. We found that $(N,M) = (4,2), (2,2),
(6,3)$ and $(3,3)$ are the only perturbatively consistent
solutions. For these models we gave 
the solutions to the tadpole cancellation conditions and the massless
spectra. In addition to the universal supergravity fields there are
various gauge and matter fields living on D-branes intersecting at
angles. The smallest intersection angle is given by $\pi/max\left( N,
  M\right)$. Explicit results are presented only for the cases where
the gauge groups are maximal, i.e.\ all the D-branes sit at the
corresponding orientifold fixed planes. These gauge groups can be
Higgsed to smaller groups by moving certain numbers of D-branes off the
O-planes. 

The type IIA orientifolds considered here can be dualized to type
IIB orientifolds by performing T-duality in the directions where ${\cal
  R}$ acts in a non-trivial way. As stated in the end of section
\ref{zweikreuzzwei} the resulting type IIB orientifolds have in most cases
a non-trivial discrete $B$ field background. (With $B$ we denote the
NSNS antisymmetric tensor.) Constant $B$ field backgrounds have
received some attention in the recent past because they can lead to a
microscopic description of non-commutative field
theories[\ref{nonco1}--\ref{noncol}]. In this context it may be also
interesting whether the IIA orientifolds considered here can be
modified to include non trivial $B$ field backgrounds. In order to
study this question one should investigate whether the tadpole
cancellation conditions can be solved by projective representations of
the orientifold group which are not equivalent to the ones given
here\cite{Douglas:1998xa,Douglas:1999hq}. Note that $B_{ij}$ is
quantized only if ${\mathcal R}$ acts with the same sign on $x^{i}$
and $x^{j}$. Otherwise $B_{ij}$ is a modulus and instead $G_{ij}$ (the
off-diagonal component of the metric) is quantized\cite{Angelantonj:2000xf}.
  
\vskip 1cm  
  
\noindent {\bf Acknowledgments} 
 
\noindent This work has been supported by TMR programs
ERBFMRX--CT96--0045 and CT96--0090.
We acknowledge discussions with Ralph Blumenhagen, Jan Conrad, Boris
K\"ors and Hans Peter Nilles.

\newpage
\begin{appendix}
\section{Computation of one-loop diagrams}\label{appa}
In this appendix we will give the details of the computation of the
diagrams fig. \ref{diagrams} in the loop channel, i.e.\ evaluate the
expressions (\ref{kb}), (\ref{an}), and (\ref{ms}). We follow closely
the notation of \cite{Blumenhagen:2000ev}. First, we introduce
abbreviations by identifying these expressions with
\begin{eqnarray}
{\cal K} & = & \left(1-1\right) 4c\int_0^\infty \frac{dt}{t^3}\left(
    \frac{1}{4NM}\sum_{n_1,k_1 
    =0}^N\sum_{n_2,k_2 =0}^M {\cal
    K}^{\left(n_1,k_1\right)\left(n_2,k_2\right)} {\cal L}_{\cal
    K}^{\left(n_1,k_1\right)\left(n_2,k_2\right)} \right) , \\
{\cal M} & = & -\left(1-1\right) c\int_0 ^\infty
    \frac{dt}{t^3}\left(\frac{1}{4NM}\sum_{n_1,k_1 =
    0}^{N}\sum_{n_2,k_2=0}^M
    \sum_{(i_1,i_2)=(0,0)}^{(N-1,M-1)}\mbox{tr}\left(
    \left(\gamma_{\Omega{\cal R}k_1k_2}^{\left(i_1, i_2
    \right)}\right)^{-1}\left(\gamma_{\Omega{\cal R}k_1k_2}^{\left(i_1-n_1,
    i_2-n_2\right)}\right) ^{T}\right)\right.\nonumber\\
& & \;\; \;\;\;\;\left. {\cal
    M}^{\left(n_1,k_1\right)\left(n_2,k_2\right)}{\cal L}_{\cal
    M}^{(n_1,k_1)(n_2,k_2)(i_1,i_2)}\right), \\
{\cal A} & = & \left(1-1\right)c\int_0^\infty
    \frac{dt}{t^3}\left(\frac{1}{4NM}\sum_{n_1,k_1 = 
    0}^{N}\sum_{n_2,k_2=0}^M
    \sum_{(i_1,i_2)=(0,0)}^{(N-1,M-1)}\mbox{tr}
    \left(\gamma_{k_1k_2}^{\left(i_1, i_2\right)}\right)
    \mbox{tr}\left(\left(\gamma_{k_1k_2}^{\left(i_1-n_1, 
    i_2-n_2\right)}\right) ^{-1}\right)\right.\nonumber\\
&& \left. \;\;\;\;\;\; {\cal 
    A}^{\left(n_1,k_1\right)\left(n_2,k_2\right)}{\cal L}_{\cal
    A}^{(n_1,k_1)(n_2,k_2)(i_1,i_2)}\right).
\end{eqnarray}
Let us first explain the meaning of the various symbols in words and
    later on give the explicit expressions. The letters ${\cal
    K}^{(\cdot )(\cdot)}$, ${\cal
    M}^{(\cdot ) (\cdot)}$ and ${\cal A}^{(\cdot )(\cdot)}$ stand for
    oscillator contributions. The upper quadruple index describes the
    twist sector and the insertion of a $\Z_N \times Z_M$ element as
    follows: the contribution corresponds to the
    $\Theta_1^{n_1}\Theta_2^{n_2}$ twisted sector with a
    $\Theta_1^{k_1}\Theta_{2}^{k_2}$ insertion in the trace. (For
    simplicity, open strings ending on different types of 
D-branes are called twisted as stated in the text.) The
    $\gamma$'s are the matrix representations of the orientifold group
    as in \cite{Gimon:1996rq}. The lower index stands for the
    corresponding group element, e.g.\ $\Omega {\cal R}k_1k_2$
    corresponds to $\Omega {\cal
    R}\Theta_1^{k_1}\Theta_{2}^{k_2}$. The upper double index
    $(i_1,i_2)$ labels the different types of D-6-branes as described
    in the text. Finally, ${\cal L}$ stands for the lattice
    contribution (i.e.\ sums over 
    discrete momenta and windings). The indexing is like in the
    oscillator contributions given above.
\subsection{Lattice contributions}

The explicit expressions for the lattice contributions are
\begin{eqnarray}
{\cal L}_{\cal K}^{\left(n_1,k_1\right)\left(n_2,k_2\right)} & = &
\chi_{\cal K} ^{\left(n_1, k_1\right)\left(n_2,k_2\right)}\mbox{Tr}^{\left(n_1
  ,n_2\right)}_{\mbox{\scriptsize KK}+\mbox{\scriptsize
  W}}\left(\Omega {\cal 
    R} \Theta_1 ^{k_1}\Theta_2 ^{k_2} e^{-2\pi t\left( L_0
      +\bar{L}_0\right)} \right), \\
{\cal L}_{\cal M}^{\left(n_1, k_1\right)\left( n_2,
    k_2\right)\left(i_1,i_2\right) }& = &\chi_{\cal M}^{\left(n_1,
    k_1\right)\left( n_2, 
    k_2\right)\left(i_1,i_2\right) 
    }\mbox{Tr}^{\left(i_1,i_2\right),\left(i_1 - n_1, i_2 -
    n_2\right)}_{\mbox{\scriptsize KK}+\mbox{\scriptsize W}} \left(
    \Omega {\cal R} \Theta_1 ^{k_1}\Theta_2 ^{k_2} e^{-2\pi t
    L_0}\right), \\
{\cal L}_{\cal A}^{\left(n_1, k_1\right)\left( n_2,
    k_2\right)\left(i_1,i_2\right) }& = &\chi_{\cal A}^{\left(n_1,
    k_1\right)\left( n_2, 
    k_2\right)\left(i_1,i_2\right) 
    }\mbox{Tr}^{\left(i_1,i_2\right),\left(i_1 -n_1, i_2 -
    n_2\right)}_{\mbox{\scriptsize KK}+\mbox{\scriptsize W}} \left(
    \Theta_1 ^{k_1}\Theta_2 ^{k_2} e^{-2\pi t
    L_0}\right).
\end{eqnarray}
In the Klein bottle $\chi $ is the number of the corresponding fixed
points\footnote{In more detail: It is the number of
  $\Theta_1^{n_1}\Theta_2^{n_2}$ fixed points which are left invariant
  under ${\cal R}\Theta_1^{k_1}\Theta_2^{k_2}$.} whereas in the open
string 
amplitudes $\chi$ is the 
intersection number of the branes involved. (The indexing is analogous
to the one described above.) The upper index at Traces gives the twist
sector.  Sums over windings and momenta lead to expressions of the
form
\begin{equation}
{\cal L}\left[ \alpha , \beta\right] \equiv \left( \sum_{m \in \Z}
  e^{-\alpha \pi t m^2 / \rho}\right)\left( \sum_{n \in \Z}e^{-\beta
  \pi t n^2 \rho}\right) ,
\end{equation}
where $\rho = r^2 / \alpha^\prime$.
In the tree channel the corresponding function is defined as
\begin{equation}
\tilde{{\cal L}}\left[ \alpha , \beta\right] \equiv \left( \sum_{m \in \Z}
  e^{-\alpha \pi l m^2  \rho}\right)\left( \sum_{n \in \Z}e^{-\beta
  \pi l n^2 /\rho}\right) .
\end{equation}
The transformation from the loop channel to the tree channel is
performed by Poisson resummation,
\begin{equation}
\sum_{n\in \Z} e^{-\pi n^2 /t} = \sqrt{t}\sum_{n\in \Z} e^{-\pi n^2 t} .
\end{equation}
The exact form of lattice contributions depends on the model and the results
are given in the text.

\subsection{Oscillator contributions}
The general expressions for the oscillator contributions are
\begin{eqnarray}
{\cal K}^{\left(n_1,k_1\right)\left(n_2,k_2\right)} & = & 
\mbox{Tr}^{\left(n_1 ,n_2\right)}_{\mbox{\scriptsize NSNS}}\left(\Omega {\cal
        R} \Theta_1^{k_1}\Theta_2^{k_2}\left(-1\right)^F e^{-2\pi
        t\left( L_0 +\bar{L}_0\right)}\right) , \label{kbo}\\
{\cal M}^{\left(n_1,k_1\right)\left(n_2,
    k_2\right)}& = & \mbox{Tr}_{\mbox{\scriptsize
    R}}^{\left(0,0\right) \left(-n_1,- 
  n_2\right)}\left(\Omega {\cal R}\Theta_1 ^{k_1}\Theta_2 ^{k_2}
e^{-2\pi tL_0}\right) , \\
{\cal A}^{\left( n_1 , k_1\right)\left(n_2 ,k_2\right) } & = &
  \mbox{Tr}^{\left( 0 , 0\right) \left(- n_1 , -
      n_2\right)}_{\mbox{\scriptsize NS}}\left( \Theta_1
    ^{k_1}\Theta_2 ^{k_2}\left(-1\right)^Fe^{-2\pi t
        L_0}\right) .
\end{eqnarray}
Because of the $\Omega {\cal R}$ insertion in the Klein bottle
partition function the expression (\ref{kbo}) is actually independent
of $k_1 , k_2$ and we define
\begin{equation}
{\cal K}^{\left( n_1 ,n_2\right)}\equiv {\cal K}^{\left(n_1
    ,k_1\right)\left( n_2 ,k_2\right)}.
\end{equation}
The upper index at the open string amplitudes indicates the boundary
conditions, i.e.\ the computation is done for a string stretching
between brane $\left( 0,0\right)$ and $\left( -n_1
  ,-n_2\right)$. Equivalently, one could trace over open strings
stretching between branes $\left( i_1 , i_2\right)$ and $\left( i_1 -
  n_1 , i_2 - n_2\right)$.

The oscillator contributions can be expressed in terms of Jacobi theta
functions and the Dedekind eta function,
\begin{eqnarray}
\vartheta\left[\begin{array}{c} \alpha \\ \beta\end{array}\right] \left(
  t\right) & 
  = & \sum_{n\in \Z} q^{\frac{\left( n+\alpha\right)^2}{2}}e^{2\pi
  i\left( n+\alpha\right)\beta}, \\
\eta\left( t\right) & = & q^{\frac{1}{24}}\prod_{n=1}^{\infty} \left(
  1 - q^n\right) ,
\end{eqnarray}
with $q = e^{-2\pi t}$. One finds
\begin{eqnarray}
{\cal K}^{\left( n_1 , n_2\right)} & = & \frac{\vartheta\left[
      \begin{array}{c} 0 \\
        1/2\end{array}\right]}{\eta^3}\prod_{n_1v_i +
    n_2w_i\not{\in}\Z}\left( \frac{ 
\vartheta\left[
      \begin{array}{c} n_1v_i + n_2w_i \\
        1/2\end{array}\right]}{\vartheta\left[
      \begin{array}{c} 
        1/2+n_1v_i +n_2w_i\\ 1/2\end{array}\right]} e^{\pi i\langle
        n_1v_i + n_2w_i\rangle}\right)\nonumber\\ & &
        \;\;\;\;\;\;\;\;\;\;\;\;\;\; 
\prod_{n_1v_i +n_2w_i\in \Z}
\left(\frac{\vartheta\left[
      \begin{array}{c} 0 \\
        1/2\end{array}\right]}{\eta^3}\right) ,\\
{\cal M}^{\left(n_1 ,k_1\right)\left(n_2 ,k_2\right)} & = &
\frac{\vartheta\left[
      \begin{array}{c} 1/2 \\
        0\end{array}\right]}{\eta^3}\prod_{\left(n_1v_i +
    n_2w_i,k_1v_i +k_2w_i\right)\not{\in}\Z^2}\left( \frac{\left(
    -2i\right)^\delta  
\vartheta\left[
      \begin{array}{c} 1/2+n_1v_i + n_2w_i \\
        k_1v_i+k_2w_i\end{array}\right]}{\vartheta\left[
      \begin{array}{c} 
        1/2+n_1v_i +n_2w_i\\1/2+k_1v_i +k_2w_i \end{array}\right]}
        e^{\pi i\langle 
        n_1v_i + n_2w_i\rangle}\right)\nonumber\\ & &
        \;\;\;\;\;\;\;\;\;\;\;\;\;\; 
\prod_{\left(n_1v_i +n_2w_i,k_1v_i +k_2w_i\right)\in \Z^2}
\left(\frac{\vartheta\left[
      \begin{array}{c} 1/2 \\
        0\end{array}\right]}{\eta^3}\right) ,\\
{\cal A}^{\left( n_1 ,k_1\right)\left(n_2 ,k_2\right)} & = &
\frac{\vartheta\left[
      \begin{array}{c} 0 \\
        1/2\end{array}\right]}{\eta^3}\prod_{\left(n_1v_i +
    n_2w_i,k_1v_i +k_2w_i\right)\not{\in}\Z^2}\left( \frac{\left(
    -2i\right)^\delta  
\vartheta\left[
      \begin{array}{c} n_1v_i + n_2w_i \\
        1/2 + k_1v_i+k_2w_i\end{array}\right]}{\vartheta\left[
      \begin{array}{c} 
        1/2+n_1v_i +n_2w_i\\1/2+k_1v_i +k_2w_i \end{array}\right]}
        e^{\pi i\langle 
        n_1v_i + n_2w_i\rangle}\right)\nonumber\\ & &
        \;\;\;\;\;\;\;\;\;\;\;\;\;\; 
\prod_{\left(n_1v_i +n_2w_i,k_1v_i +k_2w_i\right)\in \Z^2}
\left(\frac{\vartheta\left[
      \begin{array}{c} 0 \\
        1/2\end{array}\right]}{\eta^3}\right) .
\end{eqnarray}
The arguments in the theta and eta functions are $2t$ in the Klein
bottle, $t +\frac{i}{2}$ in the M\"obius strip, and $t$ in the annulus.
Further, we used the notation\cite{Blumenhagen:2000ev},
\begin{equation}
\langle x\rangle \equiv x - \left[ x\right] - \frac{1}{2} ,
\end{equation}
where the brackets on the rhs denote the integer part and
\begin{equation}
\delta =\left\{ \begin{array}{c c c}
1 & \mbox{if} & \left(n_1v_i + n_2w_i, k_1 v_i + k_2 w_i\right) \in \Z\times
  \Z+\frac{1}{2} \\
0 & \mbox{else} & \end{array}\right.
\end{equation}
The tree channel expressions $\tilde{\cal K}^{(\cdot )}$,
$\tilde{M}^{(\cdot )}$ and $\tilde{A}^{(\cdot )}$ can be evaluated with
the help of the modular transformation properties,
\begin{eqnarray}
\vartheta\left[ \begin{array}{c} \alpha \\ \beta\end{array}\right] \left(
  1/t\right) 
  & = & \sqrt{t}e^{2\pi i \alpha\beta}\vartheta \left[\begin{array}{c}
-\beta \\ \alpha\end{array}\right] \left( t\right) , \\
\eta\left( 1/t\right) & = &\sqrt{ t}\eta\left( t\right) .
\end{eqnarray}
As usual, there is a subtlety in the M\"obius strip. Before
performing the modular transformation one writes the theta functions
with complex arguments as a product of theta functions with real
arguments\cite{Gimon:1996rq}.  Since the calculation is a
straightforward modification of the one presented in the appendix
of\cite{Blumenhagen:2000ev} and the formulas are rather lengthy, we do
not give the explicit tree channel expressions here. 

\section{Tables of massless spectra}\label{spectra}

In this appendix we collect tables giving the massless spectra of 
$\Z_2\times \Z_2$, $\Z_6\times \Z_3$ and $\Z_3\times \Z_3$
$\OR$-orientifolds. The spectrum of the $\Z_4 \times \Z_2$ model is given in
the text.

\subsection{The $\Z_2\times \Z_2$ model}
\renewcommand{\arraystretch}{1.2}
  \begin{center}
    \begin{equation} \label{z2z2_closed}
      \begin{array}{|c||c|c|c|c|} \hline
        \multicolumn{5}{|c|}{\rule[-3mm]{0mm}{8mm} \text{\bf
            Closed spectrum of $\Z_2 \times \Z_2$}} \\ \hline\hline
        \text{twist-sector} & \text{\bf AAA} & \text{\bf AAB} &
        \text{\bf ABB} &
        \text{\bf BBB}\\ \hline\hline
        \mbox{untwisted} &  \multicolumn{4}{|c|}{
            \mbox{SUGRA} + 6\mbox{C}}\\ \hline
        \theta_1 & 16\mbox{C} & 16\mbox{C} & 12\mbox{C} + 4\mbox{V} &
          10\mbox{C} + 6\mbox{V}\\ \hline
        \theta_2 & 16\mbox{C} & 12\mbox{C} + 4\mbox{V} & 10\mbox{C}
          +6\mbox{V} & 10\mbox{C} +6\mbox{V}\\ \hline
        \theta_1\theta_2 & 16\mbox{C} & 12\mbox{C} + 4\mbox{V} &
          12\mbox{C} + 4\mbox{V} & 10\mbox{C}
            +6\mbox{V}\\ \hline
      \end{array}
    \end{equation}
  \end{center}
%
%
\renewcommand{\arraystretch}{1.2}
  \begin{center}
    \begin{equation}\label{z2z2_open}
      \begin{array}{|c||c|c|c|c|c|} \hline
        \multicolumn{6}{|c|}{\rule[-3mm]{0mm}{8mm} \text{\bf
            Open spectrum of $\Z_2 \times \Z_2$}} \\ \hline\hline
        \text{twist-sector} & \text{\bf AAA} & \text{\bf AAB} &
        \text{\bf ABB} & \text{\bf BBB} 
        & \mbox{gauge group / matter} \\ \hline\hline
        \begin{array}{c} \mbox{untwisted} \\ \ \\ \ \end{array}
        & \multicolumn{4}{|c|}{ 
             \begin{array}{c} 1\mbox{V} \\ 3\mbox{C} \\ \ \end{array}
             } 
        & \begin{array}{c} [Sp(\frac{M}{4})]^4 \\ (A,1,1,1) \oplus
           (1,A,1,1) \\ \oplus (1,1,A,1) \oplus (1,1,1,A) \end{array}
           \\ \hline
         \theta_1 & 1\mbox{C} & 1\mbox{C} & 2\mbox{C} & 4\mbox{C} &
        (F,F,1,1)\oplus(1,1,F,F) \\ \hline
        \theta_2 & 1\mbox{C} & 2\mbox{C} & 4\mbox{C} & 4\mbox{C} &
        (F,1,F,1)\oplus(1,F,1,F) \\ \hline
        \theta_1\theta_2 & 1\mbox{C} & 2\mbox{C} & 2\mbox{C} &
        4\mbox{C} & (F,1,1,F)\oplus(1,F,F,1) \\ \hline  
      \end{array}
    \end{equation}
  \end{center}
%

\subsection{The $\Z_6\times \Z_3$ model}
\renewcommand{\arraystretch}{1.2}
  \begin{center}
    \begin{equation}\label{z6z3_closed}
      \begin{array}{|c||c|c|} \hline
         \multicolumn{3}{|c|}{\rule[-3mm]{0mm}{8mm} \text{\bf
            Closed spectrum of $\Z_6 \times \Z_3$}} \\ \hline\hline
        \text{twist-sector} & \text{\bf ABA} & \text{\bf ABB}  \\ \hline\hline
        \mbox{untwisted}&\multicolumn{2}{|c|}{\mbox{SUGRA}+3\mbox{C}} \\ \hline
        \theta_1 + \theta_1^5 &  2\mbox{C}  &  2\mbox{C} \\ \hline
        \theta_1^2 + \theta_1^4 & 8\mbox{C} +2\mbox{V} & 8\mbox{C}
            +2\mbox{V}   \\ \hline 
        \theta_1^3  & 5\mbox{C} + 1\mbox{V} & 5\mbox{C} + 1\mbox{V}
            \\ \hline 
        \theta_2  + \theta_2^2 & 8\mbox{C} + 4\mbox{V} & 12\mbox{C} \\ \hline
        \theta_1\theta_2 + \theta_1^5\theta_2^2 & 2\mbox{C} +
            1\mbox{V} & 3\mbox{C}  \\ \hline 
        \theta_1^2\theta_2 + \theta_1^4\theta_2^2 & 8\mbox{C} +
            4\mbox{V} & 12\mbox{C}    \\ \hline 
        \theta_1^3\theta_2 + \theta_1^3\theta_2^2 & 4\mbox{C} +
            2\mbox{V} & 6\mbox{C}  \\ \hline 
        \theta_1^4\theta_2 + \theta_1^2\theta_2^2  & 9\mbox{C} +
            6\mbox{V} & 12\mbox{C} + 
            3\mbox{V} \\ \hline
        \theta_1^5\theta_2 + \theta_1\theta_2^2 & 4\mbox{C} +
            2\mbox{V} & 6\mbox{C}  \\ 
            \hline
      \end{array}
    \end{equation}
  \end{center}
%
%
\renewcommand{\arraystretch}{1.2}
  \begin{center}
    \begin{equation} \label{z6z3_open}
      \begin{array}{|c||c|c|c|} \hline
         \multicolumn{4}{|c|}{\rule[-3mm]{0mm}{8mm} \text{\bf
            Open spectrum of $\Z_6 \times \Z_3$}} \\ \hline\hline
        \text{twist-sector} & \text{\bf ABA} & \text{\bf ABB}
        &\text{representation of $U(2)\times U(2)$}\\
        \hline\hline
        \mbox{untwisted} & \multicolumn{2}{|c|}{1\mbox{V}}
        &({\bf 4},{\bf 1}_{\mbox{\scriptsize \bf 0}})
        \oplus({\bf 1}_{\mbox{\scriptsize \bf 0}},{\bf 4})  \\ 
        & \multicolumn{2}{|c|}{1\mbox{C}} 
        &({\bf 4},{\bf 1}_{\mbox{\scriptsize \bf 0}})
        \oplus({\bf 1}_{\mbox{\scriptsize \bf 0}},{\bf 4})\\
        & \multicolumn{2}{|c|}{4\mbox{C}}
        &({\bf 1},{\bf 1}_{\mbox{\scriptsize \bf 0}})
        \oplus({\bf 1}_{\mbox{\scriptsize \bf 0}},{\bf 1})\\ \hline
        \theta_1 + \theta_1^5 & \multicolumn{2}{|c|}{4\mbox{C}} & 
        ({\bf 2},\Bar{\bf 2})\oplus(\Bar{\bf 2},{\bf 2}) \\ \hline
        \theta_1^2 + \theta_1^4 
        & \multicolumn{2}{|c|}{2\mbox{C}} &
        ({\bf 4},{\bf 1}_{\mbox{\scriptsize \bf 0}})
        \oplus({\bf 1}_{\mbox{\scriptsize \bf 0}},{\bf 4})\\
        & \multicolumn{2}{|c|}{8\mbox{C}} &
        ({\bf 1},{\bf 1}_{\mbox{\scriptsize \bf 0}})
        \oplus({\bf 1}_{\mbox{\scriptsize \bf 0}},{\bf 1})  \\ \hline 
        \theta_1^3  & \multicolumn{2}{|c|}{4\mbox{C}} &
        ({\bf 2},\Bar{\bf 2})\oplus(\Bar{\bf 2},{\bf 2}) \\ \hline 
        \theta_2  + \theta_2^2 &  2\mbox{C}  & 6\mbox{C} &
        ({\bf 4},{\bf
          1}_{\mbox{\scriptsize \bf 0}})\oplus({\bf
          1}_{\mbox{\scriptsize \bf 0}},{\bf 4})\\
        &  4\mbox{C}  & 12\mbox{C} &
        ({\bf 1},{\bf 1}_{\mbox{\scriptsize \bf 0}})\\
        &  2\mbox{C}  & 6\mbox{C} &
        ({\bf 4},{\bf 1}_{\mbox{\scriptsize \bf 0}})  \\ \hline 
        \theta_1\theta_2 + \theta_1^5\theta_2^2 
        & 2\mbox{C}  & 6\mbox{C} &
        ({\bf 2},\Bar{\bf 2})\oplus(\Bar{\bf 2},{\bf 2})\\ \hline
        \theta_1^2\theta_2 + \theta_1^4\theta_2^2 
        & 2\mbox{C}  & 6\mbox{C} &
        ({\bf 1},{\bf
          1}_{\mbox{\scriptsize \bf 0}})\oplus({\bf 1_{\mbox{\scriptsize
              \bf 0}}},{\bf 1})\\
        &  4\mbox{C}  & 12\mbox{C} &
        ({\bf 1}_{\mbox{\scriptsize \bf 0}},{\bf 1})\\
        & 2\mbox{C}  & 6\mbox{C} &   
        ({\bf 1}_{\mbox{\scriptsize \bf 0}},{\bf 4}) \\ \hline
        \theta_1^3\theta_2 + \theta_1^3\theta_2^2 
        &  4\mbox{C}  & 12\mbox{C} &
        ({\bf 2},\Bar{\bf 2})\oplus(\Bar{\bf 2},{\bf 2}) \\ \hline
        \theta_1^4\theta_2 + \theta_1^2\theta_2^2  
        & 2\mbox{C}  & 6\mbox{C} &
        ({\bf 4},{\bf 1}_{\mbox{\scriptsize \bf 0}})\oplus({\bf
        1}_{\mbox{\scriptsize \bf 0}},{\bf 4})\\  
        &2\mbox{C}  & 6\mbox{C} &
        ({\bf 3},{\bf 1}_{\mbox{\scriptsize \bf 0}})\oplus({\bf
        1}_{\mbox{\scriptsize \bf 0}},{\bf 3}) \\ \hline 
        \theta_1^5\theta_2 + \theta_1\theta_2^2 
        &  4\mbox{C}  & 12\mbox{C} &
        ({\bf 2},\Bar{\bf 2})\oplus(\Bar{\bf 2},{\bf 2}) \\ \hline
      \end{array}
    \end{equation}
  \end{center}
%

\subsection{The $\Z_3\times \Z_3$ model}
\renewcommand{\arraystretch}{1.2}
  \begin{center}
    \begin{equation}\label{z3z3_closed}
      \begin{array}{|c||c|c|c|c|} \hline
         \multicolumn{5}{|c|}{\rule[-3mm]{0mm}{8mm} \text{\bf
            Closed spectrum of $\Z_3 \times \Z_3$}} \\ \hline\hline
        \text{twist-sector} & \text{\bf AAA} & \text{\bf AAB} &
            \text{\bf ABB} &
            \text{\bf BBB} \\ \hline\hline
            \mbox{untwisted} & \multicolumn{4}{|c|}{\mbox{SUGRA} + 3
            \mbox{C}} \\ \hline 
        \theta_1 + \theta_1^2 & 10\mbox{C} + 8\mbox{V} & 10\mbox{C} +
            8\mbox{V} & 12\mbox{C} + 6\mbox{V} & 18\mbox{C} \\ \hline 
        \theta_2  + \theta_2^2 & 10\mbox{C} + 8\mbox{V} & 12\mbox{C} +
            6\mbox{V} &  18\mbox{C} & 18\mbox{C}   \\ \hline 
        \theta_1\theta_2 + \theta_1^2\theta_2^2 & 10\mbox{C} +
            8\mbox{V} & 12\mbox{C} + 6\mbox{V} 
            & 12\mbox{C} + 6\mbox{V} &  18\mbox{C}   \\ \hline 
        \theta_1^2\theta_2 + \theta_1\theta_2^2  & 14\mbox{C} +
            13\mbox{V} & 15\mbox{C} + 
            12\mbox{V} & 18\mbox{C} + 9\mbox{V} & 27\mbox{C} 
               \\ \hline  
      \end{array}
    \end{equation}
  \end{center}
%
%
\renewcommand{\arraystretch}{1.2}
  \begin{center}
    \begin{equation} \label{z3z3_open}
      \begin{array}{|l|c|c|c|c|c|} \hline
        \multicolumn{6}{|c|}{\rule[-3mm]{0mm}{8mm} \text{\bf
             Open spectrum of $\Z_3 \times \Z_3$}} \\ \hline\hline
       \mbox{twist-sector} &\mbox{\bf AAA} &\mbox{\bf AAB} &\mbox{\bf ABB}
             &\mbox{\bf BBB} &
      \mbox{rep. of $SO(4)$} \\ \hline\hline
      \mbox{untwisted}&\multicolumn{4}{|c|}{1\mbox{V}} & {\bf 6}\\
        {}&\multicolumn{4}{|c|}{3\mbox{C}} &{\bf 6}\\\hline
      \theta_1 + \theta_1^2 
      & 2\mbox{C} & 2\mbox{C} & 6\mbox{C} & 18\mbox{C} &{\bf 6}\\\hline
      \theta_2 + \theta_2^2 
      & 2\mbox{C} & 6\mbox{C} & 18\mbox{C} & 18\mbox{C} &{\bf 6}\\\hline
      \theta_1\theta_2+\theta_1^2\theta_2^2 
      & 2\mbox{C} & 6\mbox{C} & 6\mbox{C} & 18\mbox{C} &{\bf 6}\\\hline
      \theta_1^2\theta_2 + \theta_1\theta_2^2 
      & 1\mbox{C} & 3\mbox{C} & 9\mbox{C} & 27\mbox{C} &{\bf 10}
      \\ \hline
      \end{array}
    \end{equation}
  \end{center}
%

\section{Projective representations of  \mbox{$\OR \times \Z_4 \times \Z_2$}}
\label{projective-rep_z4z2}

In this appendix we give the explicit expressions for the projective
representation used in the $\Z_4\times \Z_2$ orientifold in section
\ref{z4z2}. The projective representation of the $\Z_2\times
\Z_2$ subsector is chosen to be equivalent to the one presented
in\cite{Berkooz:1997dw}. 
$M_i$, $N_i$ and $D$ are as defined in \cite{Berkooz:1997dw},

\begin{equation}
  \begin{aligned}
    M_i&=\left\{
      \left(\begin{array}{cc} 0 & \unity \\ -\unity & 0
        \end{array}\right),
      \left(\begin{array}{cc} -i\sigma_2 & 0 \\ 0 &i\sigma_2
      \end{array}\right), 
      \left(\begin{array}{cc} 0 & i\sigma_2 \\ i\sigma_2 & 0
        \end{array}\right)
    \right\},\\
    D&=\left(\begin{array}{cc} 0 & -i\sigma_2 \\i\sigma_2  & 0
      \end{array}\right),\\
    N_i&\equiv DM_i=\left\{
      \left(\begin{array}{cc} i\sigma_2 & 0 \\ 0 & i\sigma_2
        \end{array}\right),
      \left(\begin{array}{cc} 0 & \unity \\ \unity & 0 \end{array}\right),
      \left(\begin{array}{cc} \unity & 0 \\ 0 & -\unity
        \end{array}\right)
    \right\},
  \end{aligned}
\end{equation}
where $\sigma_2=\left(\begin{array}{cc}0&-i\\i&0\end{array}\right)$
and the matrices fulfill $M_i^2=N_1^2=-D^2=-N_2^2=-N_3^2=-\unity$.
In addition we use the notation
\begin{equation}
  a=\frac{1}{2}\left(\unity - \sigma_2\right), \qquad \qquad b=-ia^T
\end{equation}
to define
\begin{equation}
  \begin{aligned}
    A &=\left( \begin{array}{cc} a & b \\ -b & a \end{array}\right),
    \qquad \qquad A^{-1}=A,\\
    B &=\left( \begin{array}{cc} -b & a \\ -a & -b \end{array}\right),
    \qquad \qquad B^{-1}=-B,\\
    C &=i\left( \begin{array}{cc} a & b \\ b & -a \end{array}\right),
    \qquad \qquad 
    C^{-1}=i\left( \begin{array}{cc} -a & b \\ b & a \end{array}\right),\\
    E &=i\left( \begin{array}{cc} b & -a \\ -a & -b \end{array}\right),
    \qquad \qquad 
    E^{-1}=i\left( \begin{array}{cc} b & a \\ a & -b \end{array}\right).\\
  \end{aligned}
\end{equation}
The latter satisfy $C^T=iE$. Furthermore we need the set of matrices 
\begin{equation}
  \begin{aligned}
    F &= \frac{1}{\sqrt{2}}\left( \begin{array}{cc} \unity & \unity \\ i & -i
    \end{array}\right) , \\ 
    G &= \frac{1}{\sqrt{2}}\left( \begin{array}{cc} \unity & \unity \\ -i & i
    \end{array}\right) , \\ 
    H &= \frac{1}{\sqrt{2}}\left( \begin{array}{cc} i & -i \\ \unity & \unity
    \end{array}\right) , \\ 
    K &= \frac{1}{\sqrt{2}}\left( \begin{array}{cc} i & -i \\ -\unity & -\unity
    \end{array}\right) , \\ 
  \end{aligned}
\end{equation}
which satisfy $F^{-T} = G$ and $H^{-T} = -K$.
The representation matrices of the orientifold group $\OR \times \Z_4
\times \Z_2$ are listed in table~(\ref{rep-matrices_OR_z4z2}). In the
first column, we list the systematic numbering of the branes as
explained in section~\ref{z4z2_an}, and in
the second column we give the simplified numbering used in
section~\ref{z4z2_open}. All relevant signs are listed explicitly
whereas the others are arbitrary.

The representation matrices of the orbifold group $\Z_4 \times \Z_2$
are obtained as explained in equation~(\ref{determine_gamma_Theta}) of
section~\ref{z4z2_open}. They are listed in
table~(\ref{rep-matrices_z4z2}) up to an irrelevant phase.

  \begin{center}
    \begin{equation}\label{rep-matrices_OR_z4z2}
      \begin{array}{|c|c||c|c|c|c|c|c|c|c|} \hline
        \multicolumn{10}{|c|}{\rule[-3mm]{0mm}{8mm}
          \text{\bf Representation-matrices of $\Omega \mathcal{R}
          \times \Z_4 \times \Z_2$ }}
        \\ \hline\hline
%
%
 \multicolumn{2}{|c|}{\text{brane}} & \gamma_{\Omega \mathcal{R}} &
 \gamma_{\Omega  \mathcal{R}1} & 
 \gamma_{\Omega \mathcal{R}1^2} & \gamma_{\Omega \mathcal{R}1^3} &
 \gamma_{\Omega \mathcal{R}2} & \gamma_{\Omega \mathcal{R}12} &
 \gamma_{\Omega \mathcal{R}1^22} & \gamma_{\Omega \mathcal{R}1^32}
 \\\hline\hline  
%
%
(0,0)& 1 & \unity & +A     & M_1  & +B   & M_2  & +C   & M_3  & +E\\
(1,0)& 2 & +B     & \unity & +A   & M_1  & +E   & M_2  & +C   & M_3 \\
(2,0)& 3 & N_1    & +A^T   & D    & +B^T & M_2  & -C^T & M_3  & +E^T\\
(3,0)& 4 & +B^T   & N_1    & +A^T & D    & +E^T & M_2  & -C^T & M_3\\
(0,1)& 5 & N_1    & +F     & M_3  & +G   & D    & +H   & M_2  & +K\\
(1,1)& 6 & +G     & N_1    & +F   & M_3  & +K   & D    & +H   & M_2\\
(2,1)& 7 & N_1    & -F^T   & M_2  & +G^T & M_3  & +H^T & D    & +K^T\\
(3,1)& 8 & +G^T   & N_1    & -F^T & M_2  & +K^T & M_3  & +H^T & D\\\hline 
      \end{array}
    \end{equation}
  \end{center}

  \begin{center}
    \begin{equation}\label{rep-matrices_z4z2}
      \begin{array}{|c|c||c|r|c|r|c|c|c|} \hline
        \multicolumn{9}{|c|}{\rule[-3mm]{0mm}{8mm}
          \text{\bf  Representation-matrices of $\Z_4 \times \Z_2$ }}
        \\ \hline\hline
%
%
  \multicolumn{2}{|c|}{\text{brane}} & \gamma_{1} & \gamma_{1^2} & 
\gamma_{1^3} 
& \gamma_{2} & \gamma_{12} &
 \gamma_{1^22} & \gamma_{1^32} \\\hline\hline  
%
%
(0,0)& 1 & B^{-T}    & M_1 & A^{-T}    & M_2 & E^{-T}    & M_3 & C^{-T}\\
(1,0)& 2 & B^{-T}    & M_1 & A^{-T}    & M_2 & E^{-T}    & M_3 & C^{-T}\\
(2,0)& 3 & A^{T}     & M_1 & B^{T}     & N_3 & C^{T}     & N_2 & E^{T}\\
(3,0)& 4 & A^{T}     & M_1 & B^{T}     & N_3 & C^{T}     & N_2 & E^{T}\\
(0,1)& 5 & G^{-T}N_1 & N_2 & F^{-T}N_1 & M_1 & K^{-T}N_1 & N_3 & H^{-T}N_1\\
(1,1)& 6 & G^{-T}N_1 & N_2 & F^{-T}N_1 & M_1 & K^{-T}N_1 & N_3 & H^{-T}N_1\\
(2,1)& 7 & N_1 F^T   & N_3 & N_1 G^T   & N_2 & N_1 H^T   & M_1 & N_1 K^T \\
(3,1)& 8 & N_1 F^T   & N_3 & N_1 G^T   & N_2 & N_1 H^T   & M_1 & N_1
K^T\\ \hline  
      \end{array}
    \end{equation}
  \end{center}

\end{appendix}



\begin{thebibliography}{42}
\bibitem{Polchinski:1995mt}
J.~Polchinski,
Phys.\ Rev.\ Lett.\  {\bf 75} (1995) 4724
[hep-th/9510017].
%
\bibitem{Polchinski:1996df}
J.~Polchinski and E.~Witten,
Nucl.\ Phys.\  {\bf B460} (1996) 525
[hep-th/9510169].
%
\bibitem{Sagnotti:1987tw} \label{ago1}
A.~Sagnotti,
ROM2F-87/25
{\it Talk presented at the Cargese Summer Institute on
  Non-Perturbative Methods in Field Theory, Cargese, Italy, Jul 16-30,
  1987}. 
%
\bibitem{Pradisi:1989xd}
G.~Pradisi and A.~Sagnotti,
Phys.\ Lett.\  {\bf B216} (1989) 59.
%

\bibitem{Govaerts:1989md}
J.~Govaerts,
Phys.\ Lett.\  {\bf B220} (1989) 77.
%
\bibitem{Bianchi:1990yu}
M.~Bianchi and A.~Sagnotti,
Phys.\ Lett.\  {\bf B247} (1990) 517.
%
\bibitem{Horava:1989vt}  \label{agol}
P.~Horava,
Nucl.\ Phys.\  {\bf B327} (1989) 461.
%
\bibitem{Gimon:1996rq}
E.~G.~Gimon and J.~Polchinski,
Phys.\ Rev.\  {\bf D54} (1996) 1667
[hep-th/9601038].
%
\bibitem{Dabholkar:1996zi} \label{or1}
A.~Dabholkar and J.~Park,
Nucl.\ Phys.\  {\bf B472} (1996) 207
[hep-th/9602030].
%
\bibitem{Gimon:1996ay}
E.~G.~Gimon and C.~V.~Johnson,
Nucl.\ Phys.\  {\bf B477} (1996) 715
[hep-th/9604129].
%
\bibitem{Berkooz:1997dw}
M.~Berkooz and R.~G.~Leigh,
Nucl.\ Phys.\  {\bf B483} (1997) 187
[hep-th/9605049].
%
\bibitem{Blum:1996hs}
J.~D.~Blum and A.~Zaffaroni,
Phys.\ Lett.\  {\bf B387} (1996) 71
[hep-th/9607019].
%
\bibitem{Kakushadze:1997ku}
Z.~Kakushadze and G.~Shiu,
Phys.\ Rev.\  {\bf D56} (1997) 3686
[hep-th/9705163].
%
\bibitem{Zwart:1998aj}
G.~Zwart,
Nucl.\ Phys.\  {\bf B526} (1998) 378
[hep-th/9708040].
%
\bibitem{Forste:1998bd}
S.~F\"orste and D.~Ghoshal,
Nucl.\ Phys.\  {\bf B527} (1998) 95
[hep-th/9711039].
%
\bibitem{O'Driscoll:1998mk}
D.~O'Driscoll,
hep-th/9801114.
%
\bibitem{Aldazabal:1998mr} \label{orl}
G.~Aldazabal, A.~Font, L.~E.~Ibanez and G.~Violero,
Nucl.\ Phys.\  {\bf B536} (1998) 29
[hep-th/9804026].
%
\bibitem{Berkooz:1996km}
M.~Berkooz, M.~R.~Douglas and R.~G.~Leigh,
Nucl.\ Phys.\  {\bf B480} (1996) 265
[hep-th/9606139].
%
\bibitem{Blumenhagen:2000md}
R.~Blumenhagen, L.~G\"orlich and B.~K\"ors,
Nucl.\ Phys.\  {\bf B569} (2000) 209
[hep-th/9908130].
%
\bibitem{Blumenhagen:2000ev}
R.~Blumenhagen, L.~G\"orlich and B.~K\"ors,
JHEP {\bf 0001} (2000) 040
[hep-th/9912204].
%
\bibitem{Polchinski:1988tu}
J.~Polchinski and Y.~Cai,
Nucl.\ Phys.\  {\bf B296} (1988) 91.
\bibitem{Klein:2000tf}
M.~Klein and R.~Rabadan,
JHEP {\bf 0007} (2000) 040
[hep-th/0002103].
\bibitem{Kakushadze:1998eg}
Z.~Kakushadze,
Nucl.\ Phys.\  {\bf B535} (1998) 311
[hep-th/9806008].
\bibitem{Bianchi:1992eu}
M.~Bianchi, G.~Pradisi and A.~Sagnotti,
Nucl.\ Phys.\  {\bf B376} (1992) 365.
\bibitem{Blumenhagen:2000fp}
R.~Blumenhagen, L.~G\"orlich, B.~K\"ors and D.~L\"ust,
Nucl.\ Phys.\  {\bf B582} (2000) 44
[hep-th/0003024].
\bibitem{Lauer:1989ax}
J.~Lauer, J.~Mas and H.~P.~Nilles,
Phys.\ Lett.\  {\bf B226} (1989) 251.
\bibitem{Lauer:1991tm}
J.~Lauer, J.~Mas and H.~P.~Nilles,
Nucl.\ Phys.\  {\bf B351} (1991) 353.
%
\bibitem{Douglas:1998fm}  \label{nonco1}
M.~R.~Douglas and C.~Hull,
JHEP {\bf 9802} (1998) 008
[hep-th/9711165].
%
\bibitem{Connes:1998cr}
A.~Connes, M.~R.~Douglas and A.~Schwarz,
JHEP {\bf 9802} (1998) 003
[hep-th/9711162].
\bibitem{Schomerus:1999ug}
V.~Schomerus,
JHEP {\bf 9906} (1999) 030
[hep-th/9903205].
%
\bibitem{Seiberg:1999vs}  \label{noncol}
N.~Seiberg and E.~Witten,
JHEP {\bf 9909} (1999) 032
[hep-th/9908142].
%
\bibitem{Douglas:1998xa}
M.~R.~Douglas,
hep-th/9807235.
\bibitem{Douglas:1999hq}
M.~R.~Douglas and B.~Fiol,
hep-th/9903031.
%
\bibitem{Angelantonj:2000xf}
C.~Angelantonj and R.~Blumenhagen,
Phys.\ Lett.\  {\bf B473} (2000) 86
[hep-th/9911190].
\end{thebibliography}
\end{document}